\documentclass{article}

\usepackage[main,final]{neurips_2026}

\usepackage[utf8]{inputenc}
\usepackage[T1]{fontenc}
\usepackage{microtype}
\usepackage{graphicx}
\usepackage{booktabs} % for professional tables
\usepackage{tabularx}
\usepackage{longtable}
\newcolumntype{Y}{>{\raggedright\arraybackslash}X}
\usepackage{multirow}
\usepackage{url}
\usepackage{xcolor}
\newcommand{\blue}[1]{{\color{black}#1}}
\usepackage{etoolbox}
\usepackage{float}
\usepackage{needspace}
\usepackage{placeins}

\usepackage{amsmath}
\usepackage{amssymb}
\usepackage{mathtools}
\usepackage{amsthm}
\usepackage{hyperref}
\hypersetup{hidelinks}

\usepackage[capitalize,noabbrev]{cleveref}

\makeatletter
\let\TWNMoriginalNotice\@noticestring
\renewcommand{\@noticestring}{\blue{\TWNMoriginalNotice}}
\renewcommand{\acksection}{\section*{\blue{Acknowledgments and Disclosure of Funding}}}

\let\TWNMoriginalLBibitem\@lbibitem
\renewcommand{\@lbibitem}[2][]{%
  \ifboolexpr{
    test {\ifstrequal{#2}{jiang2026sciphi}} or
    test {\ifstrequal{#2}{sakshi2025spur}} or
    test {\ifstrequal{#2}{wan-etal-2025-speechiq}} or
    test {\ifstrequal{#2}{Hu_2026_CVPR}}
  }{\color{black}}{\color{black}}%
  \TWNMoriginalLBibitem[#1]{#2}}
\makeatother

\theoremstyle{plain}

\theoremstyle{definition}

\theoremstyle{remark}

\usepackage[most]{tcolorbox}
\tcbset{colback=white,colframe=black!40,arc=2mm,boxrule=0.5pt}

\newtcolorbox{promptbox}[1][]{%
  enhanced,
  colback=purple!5!white,
  colframe=purple!60!black,
  arc=2mm,
  boxrule=0.5pt,
  left=1mm,
  right=1mm,
  top=1mm,
  bottom=1mm,
  fonttitle=\bfseries,
  title={#1}
}

\title{The World is Not Mono: Enabling Spatial Understanding in Large Audio-Language Models}

\author{
\blue{Yuhuan You \qquad Lai Wei \qquad Tianshu Qu}\\
\blue{State Key Laboratory of General Artificial Intelligence}\\
\blue{School of Intelligence Science and Technology}\\
\blue{Peking University}\\
\blue{Beijing, China}\\
\blue{\texttt{qutianshu@pku.edu.cn}}
}

\begin{document}
\raggedbottom

\maketitle

\begin{abstract}
  Large audio-language models have made rapid progress in recognizing what is present in an audio clip, yet spatial audio-language understanding still lacks a clear task interface. A model must not only identify sound events, but also decide where they occur, which semantic and spatial attributes belong to the same auditory object, how multiple objects are arranged, and whether a scene-level answer is physically plausible. We formalize this missing capability as audio scene analysis (ASA), a three-level problem spanning atomic perception, relational integration, and cognitive reasoning.
  We propose \emph{The World is Not Mono} (TWNM), a framework that instantiates this definition by equipping audio-language models with explicit spatial evidence. TWNM uses physically grounded First-Order Ambisonics (FOA) simulation to obtain controllable supervision, learns slot-regularized spatial representations from multichannel audio, and fuses these representations with semantic audio features before reasoning with a language model. We further train the model with a progressive curriculum, ending with preference optimization over metadata-derived correct answers and auxiliary format/evidence rewards. To operationalize the ASA definition, we build a controlled benchmark from scene metadata, covering localization, attribute binding, spatial comparison, scene abduction, and counterfactual reasoning.
  On this ASA benchmark, TWNM achieves 70.8\% overall accuracy, 66.4\% on spatial-family tasks, and 79.76\% on mixed $\mathcal{L}_3$ scene-level question answering (QA) under exact multiple-choice question answering (MCQA) scoring. We also audit monaural and binaural reference systems as diagnostic references with explicit audit labels, because they differ in spatial input, training interface, and output format. The supported claim is that a clearly defined ASA task hierarchy, FOA-conditioned spatial representations, and metadata-grounded training together enable controlled, auditable spatial audio-language reasoning, with STARSS23 providing a limited real-recording diagnostic.
\end{abstract}

\section{Introduction}

Large audio-language models (LALMs) can recognize speech, music, and environmental sounds, and can answer open-ended questions about semantic audio content \citep{qwen2audio2024, airbench2024, salmonn2023}. Yet most of them still process audio as a monophonic semantic stream. They focus on what is present, but provide little explicit grounding for where sources are located, how far they are, how multiple sources are arranged, or whether an answer is physically consistent with the acoustic scene. This gap matters for robotics, augmented reality, assistive listening, and embodied interaction, where an agent must reason about a three-dimensional sound field.

We argue that spatial audio-language understanding is broader than source localization. Complete audio scene analysis also requires binding semantic and spatial attributes into coherent auditory objects, comparing relations among objects, and using acoustic evidence together with physical and commonsense priors. Motivated by classical Auditory Scene Analysis (ASA) \citep{bregman1990}, we study a three-level hierarchy: atomic perception, relational integration, and cognitive reasoning.

Reliable supervision is the main obstacle. Real recordings rarely provide source identity, 3D location, distance, room geometry, reverberation, and multi-source binding annotations simultaneously, while large semantic audio datasets usually lack spatial metadata \citep{fsd50k, clotho2019, starss23}. We therefore use physically grounded First-Order Ambisonics (FOA) simulation to construct controllable training and evaluation data with verifiable source, room, and relation metadata. We further adapt and evaluate the spatial encoder on real STARSS23 recordings as a limited real-recording diagnostic beyond simulation.

We propose \emph{The World is Not Mono} (TWNM), a spatial audio-language framework that learns slot-regularized spatial representations from multichannel FOA input, fuses them with semantic audio features, and aligns a language model with supervised and preference-based training. We also build an ASA benchmark from held-out simulator metadata, covering localization, attribute binding, spatial comparison, scene abduction, and counterfactual reasoning. TWNM achieves 70.8\% overall accuracy, 66.4\% on spatial-family tasks, and 79.76\% on mixed $\mathcal{L}_3$ scene-level question answering (QA) under exact multiple-choice question answering (MCQA) scoring. External monaural and binaural systems are reported as diagnostic references with explicit audit labels, because they differ from TWNM in spatial-channel access, answer interface, and scoring protocol.

\section{Related Work}
\label{sec:relatedwork}

Recent LALMs connect acoustic encoders to large language models (LLMs) for instruction following and open-ended question answering (QA) over speech, sound, and music \citep{salmonn2023, qwen2audio2024, audioflamingo2024, audioflamingo2_2025, qwen25omni2025, goel2025audioflamingo3advancing, audioflamingonext2026}; benchmarks such as AIR-Bench evaluate broad semantic comprehension \citep{airbench2024}. These systems improve general audio understanding, but direction, distance, room acoustics, and cross-source spatial relations are usually not first-class modeling or evaluation targets.

Spatial audio-language modeling is emerging. BAT introduces spatial QA over binaural mixtures \citep{BAT2025}; ELSA learns spatially aware audio-text embeddings for retrieval and localization \citep{elsa2025}; and recent FOA-based spatial QA work builds questions from STARSS23 spatiotemporal descriptions \citep{spatialaudioqa2025}. In parallel, computational ASA and sound event localization and detection (SELD) methods predict event classes, directions, and distances from multichannel audio \citep{acccdoa2020, shimada2021acccdoa, starss23, dcase2024selddistance}. These lines provide important perceptual foundations, but they typically stop at structured labels, retrieval, or task-specific QA rather than auditable multiple-choice reasoning over bound spatial-semantic scene structure. TWNM targets this missing interface by combining explicit FOA-conditioned spatial representations, simulator-derived metadata supervision, and language-model alignment.

\blue{Sci-Phi~\citep{jiang2026sciphi} combines FOA spatial and $W$-channel spectral encoders for structured descriptions of sources, background, and room properties. SPUR~\citep{sakshi2025spur} augments a language model with an FOA encoder, adapter, and LoRA-based spatial-QA training. TWNM organizes ASA around perception, source-specific binding, and scene reasoning, with separately organized spatial and semantic representations fused for question answering. Appendix~\ref{app:spatial_system_comparison} reports the complementary within-study evidence. SpeechIQ organizes speech evaluation into memory, understanding, and application \citep{wan-etal-2025-speechiq}; its speech-content tasks complement ASA's source-specific geometry. In visual grid tasks, task-relevant positional structure also supports abstract reasoning \citep{Hu_2026_CVPR}.}

\section{Task Definition of Audio Scene Analysis}
\label{sec:problem_formulation}

\begin{figure}[t]
  \centering
  \includegraphics[width=\linewidth]{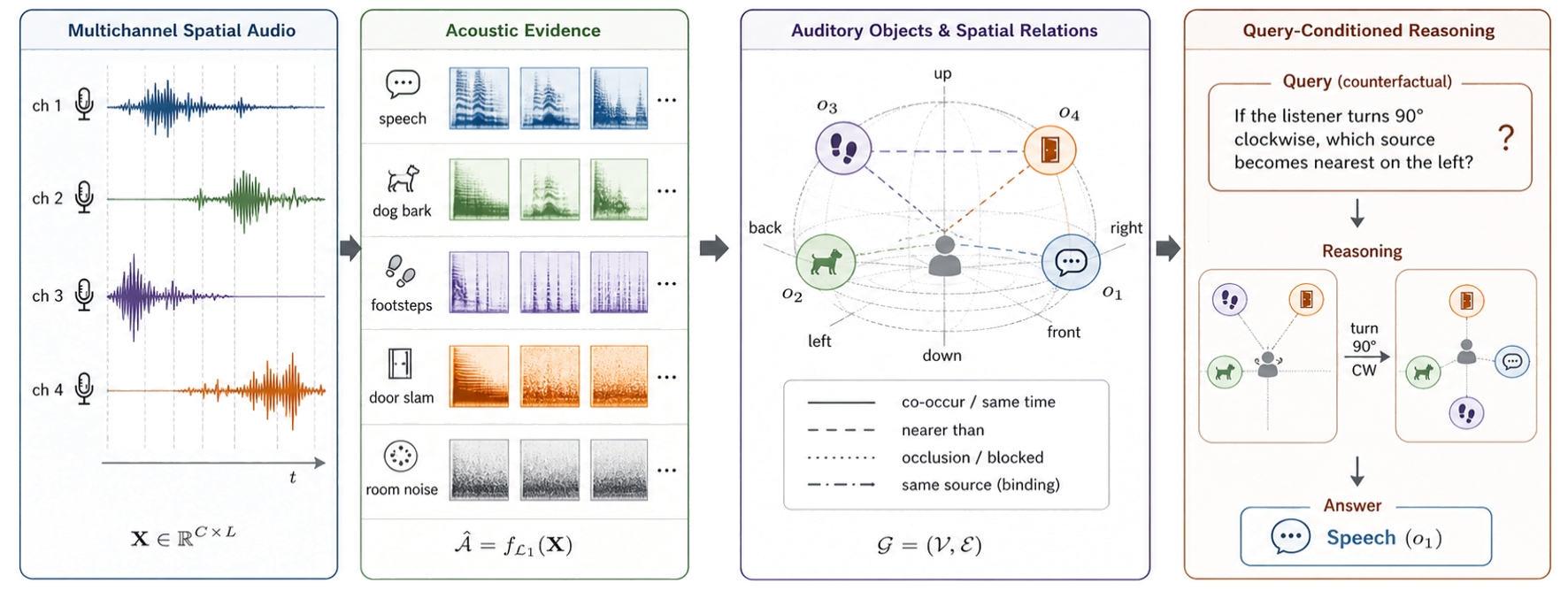}
  \caption{Three-level task hierarchy for spatial audio scene analysis. Multichannel spatial audio provides candidate acoustic evidence; relational integration binds this evidence into auditory objects and spatial relations; cognitive reasoning uses the structured scene to answer query-conditioned spatial questions.}
  \label{fig:asa_task_hierarchy}
\end{figure}

Given a multichannel waveform $\mathbf{X}$ and a text query $q$ with answer options, ASA asks for an audio-grounded answer $Y$ supported by the underlying auditory scene. This interface is intentionally compatible with multiple-choice evaluation, but the target is broader than selecting a label: the answer may depend on source identity, listener-centric geometry, environment acoustics, or relations among simultaneous sources. We organize this task through three descriptive evaluation levels rather than an implemented symbolic inference pipeline: \emph{Atomic Perception} ($\mathcal{L}_1$), \emph{Relational Integration} ($\mathcal{L}_2$), and \emph{Cognitive Reasoning} ($\mathcal{L}_3$). The hierarchy follows the transition from acoustic evidence to bound auditory objects and then to query-conditioned scene reasoning. Figure~\ref{fig:asa_task_hierarchy} gives the conceptual structure.

An auditory scene $\mathcal{S}$ is observed through a multichannel waveform $\mathbf{X}\in\mathbb{R}^{C\times L}$, where $C$ is the channel count and $L$ is the number of samples. Let $\mathcal{O}=\{o_1,\ldots,o_N\}$ denote the $N$ auditory objects in the scene, including foreground source events and background/environment objects. Each object is represented as
\begin{equation}
  o_i=(c_i,\mathbf{s}_i,\tau_i),
\end{equation}
where $c_i\in\mathcal{C}_{\mathrm{obj}}$ is a semantic class, transcript-like attribute, or environment category from the object label space $\mathcal{C}_{\mathrm{obj}}$; $\mathbf{s}_i\in\mathbb{R}^3\cup\{\varnothing\}$ is a listener-centric coordinate for localizable objects and a null coordinate for diffuse environment objects; and $\tau_i$ is the temporal support of the object within the clip. Room size, absorption, and reverberation are scene-level acoustic attributes. We use $\mathcal{K}$ for the physical priors, scene schemas, and causal regularities available to the language decoder through pretraining and alignment; it is not an external symbolic rule base.

\paragraph{$\mathcal{L}_1$: atomic perception.}
The first level extracts candidate attributes from waveform evidence. Formally,
\begin{equation}
  \hat{\mathcal{A}}=f_{\mathcal{L}_1}(\mathbf{X}),
\end{equation}
where $f_{\mathcal{L}_1}$ is the attribute extractor and $\hat{\mathcal{A}}$ contains source count, event identity, speech content, environment category, azimuth, elevation, distance, and room-acoustic evidence. These attributes are initially isolated observations: the model may estimate that an event class is present and that a source lies in a direction, but $\mathcal{L}_1$ does not require these observations to be bound to the same object. In the benchmark, this level corresponds to counting, event or speech identification, absolute direction, distance, and environment-acoustics questions.

\paragraph{$\mathcal{L}_2$: relational integration.}
The second level binds candidate attributes into a relational scene abstraction $\mathcal{G}=(\mathcal{V},\mathcal{E})$, where $\mathcal{V}$ is the set of validated object nodes and $\mathcal{E}$ is the set of relation edges. Intra-object binding asks whether observed attributes $\mathbf{a}\subset\{c,\mathbf{s},\tau\}$ belong to the same object and can be written as
\begin{equation}
  \hat{\bar{\mathbf{a}}}=\operatorname*{argmax}_{\bar{\mathbf{a}}}P(\bar{\mathbf{a}}\mid\mathbf{a},\mathbf{X}).
\end{equation}
where $\bar{\mathbf{a}}$ denotes missing complementary attributes and $P$ is the model-implied conditional distribution. This covers both location retrieval from source identity and source retrieval from spatial attributes. Inter-object relations are represented as
\begin{equation}
  r_{ij}=\Phi(o_i,o_j),\qquad i\neq j,
\end{equation}
where $r_{ij}$ is the relation between objects $o_i$ and $o_j$, and $\Phi$ is a relation operator covering relative direction, distance comparison, co-occurrence, and physical consistency under the inferred layout. The corresponding benchmark families test attribute verification, querying a location from a named source, querying a source from a location, and comparing directions or distances across objects.

\paragraph{$\mathcal{L}_3$: cognitive reasoning.}
The third level performs query-conditioned inference over acoustic evidence, the relational abstraction, and implicit knowledge:
\begin{equation}
  Y^*=\operatorname*{argmax}_Y P(Y\mid q,\mathcal{G},\mathcal{K},\mathbf{X}).
\end{equation}
Here $q$ is the text query, $Y$ is a candidate answer, and $Y^*$ is the optimal audio-grounded answer. This level includes scene abduction, causal-intent style interpretation, physical consistency, source-removal counterfactuals, observer rotation, and multi-hop spatial composition. These questions require the model to use bound acoustic evidence together with physical and scene priors, rather than merely recognizing a sound class. TWNM does not emit an explicit symbolic graph at inference time; the graph notation defines the evaluation target, while the implemented model represents objects and relations through supervised source slots and dense acoustic tokens.

\section{Methodology}
\label{sec:method}

TWNM instantiates the ASA hierarchy with a compact spatial-audio language pipeline. FOA input is represented as a 4-channel B-format signal $\mathbf{x}(t)=[W(t),Y(t),Z(t),X(t)]^T$, where $W$ is the omnidirectional pressure channel and $\{X,Y,Z\}$ are first-order pressure-gradient channels \citep{rafaely2015fundamentals}. Directional cues are available from the multichannel structure, while distance is learned from level, direct-to-reverberant structure, room acoustics, and source statistics.

\subsection{Architecture}
\label{sec:architecture}

The model uses two audio streams. The semantic branch uses the frozen \texttt{openai/whisper-small} encoder on the FOA $W$ channel to extract speech and event features \citep{radford2022whisper}. The spatial branch processes the 44.1 kHz FOA waveform with an 8-layer encoder trained to recover source slots, direction, distance, source count, and room attributes. The two streams are temporally aligned, fused by a dense hybrid projector, and inserted as continuous audio tokens into an Audio-Flamingo-3-series Qwen2.5-7B decoder \citep{goel2025audioflamingo3advancing,qwen25technical2024}.

\begin{figure}[t]
  \centering
  \includegraphics[width=0.74\columnwidth]{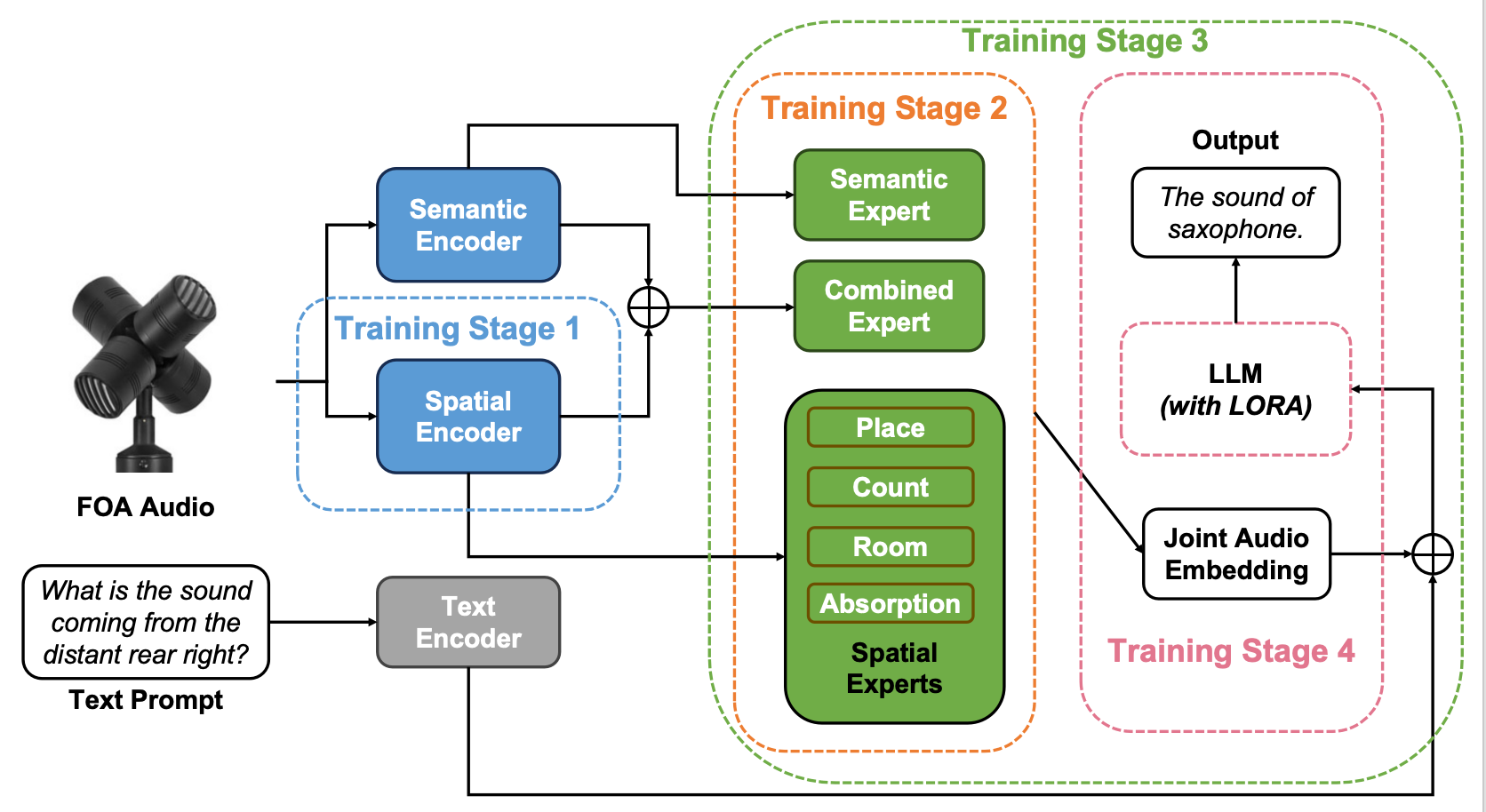}
  \caption{TWNM model architecture. A semantic branch and a spatial branch are fused through hybrid projection before language-model reasoning.}
  \label{fig:framework}
\end{figure}

\paragraph{Spatial encoder.}
\label{sec:spatial_encoder}
The spatial encoder is inspired and adapted from the envelope-separation-aided multi-task learning model for blind source counting and localization \citep{du2025envelope}. It first converts the FOA waveform into complex short-time Fourier transform (STFT) features and unfolds real and imaginary components into an 8-channel spectro-temporal map. Grouped spectral projection preserves FOA channel structure; stacked cross-band, full-band, and narrow-band blocks model frequency interaction and temporal context; and a long short-term memory (LSTM) attractor module produces up to $K=3$ source slots, where $K$ is the maximum number of modeled sources. Each active slot predicts direction, distance, event class, and existence probability, while a global token predicts room area and absorption. The dense encoder map, rather than only the pooled heads, is projected to 768 dimensions and aligned to the Whisper sequence length before fusion.

Active ground-truth sources are matched to predicted slots with Hungarian assignment over direction, distance, class, objectness, and no-object costs. The Stage-1 objective combines direction, distance, class, activity, count, and room-attribute terms:
\begin{equation}
  \mathcal{L}_{\text{spat}}
  = \lambda_{\text{dir}}\mathcal{L}_{\text{dir}}
  + \lambda_{\text{dist}}\mathcal{L}_{\text{dist}}
  + \lambda_{\text{cls}}\mathcal{L}_{\text{cls}}
  + \lambda_{\text{obj}}\mathcal{L}_{\text{obj}}
  + \lambda_{\text{cnt}}\mathcal{L}_{\text{cnt}}
  + \lambda_{\text{room}}\mathcal{L}_{\text{room}}.
\end{equation}
For STARSS23 adaptation, class and room losses are masked because the real recordings use a different event vocabulary and lack room-parameter labels; direction, distance, activity, and count supervision remain active.

\paragraph{Hybrid projection.}
\label{sec:hybrid_projector}
\blue{The dense hybrid projector computes a semantic transformation of $\mathbf{H}_{sem}$, four parallel spatial transformations of $\mathbf{H}_{spat}$, and a combined transformation of $\mathbf{H}_{sem}+\mathbf{H}_{spat}$. Concatenation keeps semantic, spatial, and combined views available before bottleneck compression into the decoder's acoustic tokens.}

Concretely, the \blue{parallel spatial pathways produce} $\mathbf{O}_{sp}=\mathrm{Concat}_{k=1}^4[\mathcal{E}_{sp}^k(\mathbf{H}_{spat})]$, and the projector outputs
\begin{equation}
  \mathbf{E}_{audio}=\mathbf{W}_2\,\mathrm{GELU}\left(\mathbf{W}_1\,\mathrm{LN}\left(
  \mathrm{Concat}[\mathcal{E}_{sem}(\mathbf{H}_{sem}),\mathbf{O}_{sp},\mathcal{E}_{comb}(\mathbf{H}_{sem}+\mathbf{H}_{spat})]\right)\right).
\end{equation}
This projected sequence is the soft-token audio interface consumed by the decoder.

\blue{The two streams organize the same mixture differently. Whisper provides open-vocabulary speech and event content, while source-wise geometric and class supervision organizes the dense spatial map. With overlapping sources, semantic evidence can be temporally diffuse while directional cues depend on channel and source correspondence. Preserving semantic, spatial, and combined views before compression gives the projector access to both kinds of evidence when constructing the decoder's acoustic tokens. The projector and attractor ablations in Appendix~\ref{app:ablation_details} evaluate this interface before the final policy stage.}

\paragraph{Decoder integration.}
\label{sec:backbone_integration}
The projected audio sequence is substituted at a reserved \texttt{<AcousticTokens>} anchor in the text prompt and processed by a standard causal decoder without adding cross-attention layers. During supervised tuning, the loss is applied only to reasoning and answer tokens; prompt tokens and inserted audio embeddings are masked. This concentrates optimization on audio-conditioned answer generation while preserving the pretrained decoder's language ability.

\subsection{Progressive Training}
\label{sec:curriculum}

Training proceeds in four stages. Stage 1 trains the spatial encoder with simulator labels, then adapts it on a 70/30 simulation-STARSS23 mixture while masking unavailable real-recording semantic and room losses. Stage 2 freezes both encoders and the decoder and aligns only the dense hybrid projector. Stage 3 jointly trains the projector and decoder low-rank adaptation (LoRA) adapters with \blue{supervised fine-tuning (SFT) on reasoning and final-answer tokens while the encoders remain frozen}. Stage 4 applies \blue{the existing} Soft Adaptive Policy Optimization (SAPO) \citep{gao2025softadaptivepolicyoptimization}, using metadata-derived correctness as the dominant reward with auxiliary format, validity, length, evidence, and hidden-reference penalties. \blue{Only decoder LoRA is trained in Stage 4; both encoders and the projector are frozen. Our contribution is the metadata-grounded task reward.} Full losses and hyperparameters are in Appendix~\ref{app:training_details}.

\blue{The simulator stores the hidden state of each rendered scene, allowing fresh completions to be scored against the corresponding metadata. This makes the task reward computable during online policy optimization. For a source-removal question, correctness follows from the remaining sources; for an observer-rotation question, it follows from the transformed listener-centric geometry. The dominant correctness term rewards the final option derived from that state. Auxiliary terms shape the response by rewarding valid answer structure and audible-evidence use, and penalizing references to hidden metadata. The response-level reward jointly shapes how the decoder uses the acoustic evidence, reasons over it, and expresses the answer. Appendix~\ref{app:training_details} specifies the reward terms and weights.}

\section{Experiments}
\label{sec:exp}

This section evaluates TWNM on the ASA benchmark and on complementary real-recording analyses. The ASA benchmark is the primary setting because simulator metadata provides exact source identity, geometry, distance, and room-acoustic ground truth.

\subsection{Data, Benchmark, and Scoring}
\label{subsec:datasets}
\label{sec:benchmark}

TWNM uses separate corpora for encoder supervision, SFT, SAPO, and held-out evaluation. The encoder is trained on a 50k-scene full-simulation corpus from FSD50K \citep{fsd50k}. The LALM stages use separate 50k-scene SFT and SAPO corpora rendered from Clotho \citep{clotho2019}, LibriTTS-R \citep{koizumi2023librittsrrestoredmultispeakertexttospeech}, and SongDescriber \citep{manco2023songdescriberdatasetcorpus} with a 5:3:2 event/speech/music mixture. The final 1,000-question ASA benchmark is selected from an independently rendered 10k-scene benchmark corpus drawn from the same public source datasets; its rendered scenes, spatial configurations, rich text scene descriptions (RTSDs), questions, and answer keys are excluded from SFT and SAPO batches. The benchmark is therefore a held-out spatial-composition and QA test, not an unseen dry-clip identity test, because dry-source clip identities can overlap across render corpora.

\blue{A dry clip's identity alone does not determine the spatial relations queried by ASA. Left/right placement, near/far distance, co-source combinations, and counterfactual answers depend on the new room draw, source positions, and mixture. Rendering that composition produces a new FOA waveform and the metadata from which its questions and answers are derived. Reusing a bark recording, for example, leaves its location and relation to simultaneous speech undetermined. The benchmark therefore tests whether the model can recover and bind physical evidence in held-out rendered compositions and use it to answer held-out questions. Of 10,154 unique dry clips in the benchmark render corpus, 5,911 also occur in SFT and 5,967 in SAPO.}

Each scene is rendered in a shoebox room with \texttt{pyroomacoustics} \citep{pyroomacoustics2017}, a four-channel FOA receiver, and up to three spatialized sources. The simulator stores source identity, position, distance, room geometry, absorption, and reverberation metadata. Gemini 3 Flash converts hidden RTSD metadata into audio-only multiple-choice questions (MCQs) with four choices and a single gold answer letter; the model sees only the audio and question/options at evaluation time. The teacher controls language surface and distractors, but correctness remains tied to simulator metadata rather than to teacher preference. The benchmark contains 385 $\mathcal{L}_1$, 279 $\mathcal{L}_2$, and 336 $\mathcal{L}_3$ questions. Appendix~\ref{app:data_pipeline} gives the rendering and holdout details; Appendices~\ref{app:prompt} and~\ref{app:benchmark_taxonomy} provide prompt contracts, reader-facing examples, and the full task taxonomy, while the complete task-family results are reported in Table~\ref{tab:main_task_family_results}.
For example, a held-out item can ask which source would move to the listener's left after a $180^\circ$ observer rotation; the gold option is computed from simulator coordinates rather than inferred by the teacher LLM.

\blue{Metadata verification and human validation address different aspects of benchmark quality. For source-removal and observer-rotation questions, the gold answer follows from the stored scene state and transformation rule. Metadata comparison checks this physical consistency. Human validation tests whether the resulting wording is understandable, admits a unique supported answer, and agrees with the audible content. Two annotators independently judged 180 frozen questions, with 10 from each of the 18 families, before seeing the hidden gold answers. An independent adjudicator reviewed 36 flagged cases. After adjudication, selected answers agreed with hidden gold answers on 172 of the 180 questions (95.6\%; Wilson 95\% CI $[91.5,97.7]$). These measurements describe the validation subset (Appendix~\ref{app:human_validation}).}

\subsection{Spatial Encoder Robustness}
\label{sec:encoder_analysis}

Before end-to-end QA, we evaluate whether the Stage-1 encoder \blue{recovers the physical quantities targeted by its auxiliary heads}. Table~\ref{tab:encoder_diagnostics} compares EINV2 \citep{nakashima2021einv2}, Sp-ACCDOA \citep{shimada2021acccdoa}, and TWNM's encoder on source count, direction-of-arrival (DOA), and distance under the same retained-segment and Hungarian-matching evaluation.

\Needspace{5\baselineskip}
STARSS23 evaluation uses static one-second development-evaluation segments after excluding empty, time-varying, and more-than-three-source clips. TWNM has the best diagnostic numbers in simulation and after STARSS23 sim-to-real adaptation \citep{starss23}. We use this as evidence that the acoustic-token interface can recover useful physical variables, while Appendix~\ref{app:spatial_encoder_results} gives the filtering and matching protocol and Appendix~\ref{app:seld_threshold} reports the SELD threshold sensitivity needed to interpret source-count tokenization.

\blue{Before fusion with Whisper, the frozen spatial encoder already carries source-specific semantic information. We evaluate it on 1,000 held-out full-simulation scenes containing 1,493 sources. Hungarian assignment matches predicted slots to reference sources using direction, distance, and objectness, with class cost set to zero. Under this assignment, slot-level class predictions are correct for 666 of 1,493 sources (44.61\%); DOA error falls below $20^\circ$ for 90.56\%; and 41.39\% satisfy both criteria. Across 404 multi-source scenes, top-1 class predictions match their spatially assigned source at 37.90\%, compared with 24.23\% across other-source comparisons within the same mixture. The binding advantage is 13.68 percentage points (scene-bootstrap 95\% CI $[10.64,16.77]$; Appendix~\ref{app:binding_diagnostic}). These results associate semantic identity with the corresponding geometry before fusion. Whisper subsequently contributes richer open-vocabulary speech and event content.}

\begin{table}[t]
  \centering
  \caption{Stage-1 spatial encoder results on retained synthetic segments and retained static STARSS23 one-second development-evaluation segments. Count accuracy is reported in percent; DOA and distance are mean absolute errors (MAEs).}
  \label{tab:encoder_diagnostics}
  {\small
    \begin{tabular}{@{}llccc@{}}
      \toprule
      \textbf{Domain} & \textbf{Method} & \textbf{Count Acc. (\%)} & \textbf{DOA MAE ($^{\circ}$)} & \textbf{Dist. MAE (m)} \\
      \midrule
      \multirow{3}{*}{Simulation} & EINV2 & 76.8 & 30.4 & 0.642 \\
      & Sp-ACCDOA & 64.9 & 26.0 & 0.506 \\
      & \textbf{TWNM Encoder} & \textbf{89.4} & \textbf{7.9} & \textbf{0.270} \\
      \midrule
      \multirow{3}{*}{STARSS23} & EINV2 & 58.5 & 35.6 & 0.399 \\
      & Sp-ACCDOA & 54.1 & 33.0 & 0.434 \\
      & \textbf{TWNM Encoder (Sim-to-Real)} & \textbf{81.7} & \textbf{22.5} & \textbf{0.380} \\
      \bottomrule
    \end{tabular}%
  }
\end{table}

\Needspace{5\baselineskip}
\subsection{Main Results}
\label{sec:main_results}

\begin{table}[!htbp]
  \centering
  \caption{Representative ASA benchmark diagnostics. Values are audited four-choice QA correctness percentages. External rows are diagnostic references rather than strict same-input baselines. Audit labels identify non-equivalent scoring protocols; ``Exact avg.'' denotes the mean of original-order and answer-shuffled exact-option audits. Monaural baselines receive the FOA W channel, BAT-SFT receives matched binaural audio, and TWNM receives full FOA. Full audits are in Appendix~\ref{app:additional_baselines}.}
  \label{tab:main_results_transposed}
  \small
    \begin{tabularx}{\linewidth}{@{}>{\raggedright\arraybackslash}Xllcccc@{}}
      \toprule
      \multirow{2}{*}{\textbf{Model / Stage}} & \multirow{2}{*}{\textbf{Input}} & \multirow{2}{*}{\textbf{Audit}} & \multicolumn{3}{c}{\textbf{ASA Competency Levels}} & \multirow{2}{*}{\textbf{Overall (\%)}} \\
      \cmidrule(lr){4-6}
      & & & $\mathcal{L}_1$ & $\mathcal{L}_2$ & $\mathcal{L}_3$ & \\
      \midrule
      Random & -- & Analytic & 25.00 & 25.00 & 25.00 & 25.00 \\
      BAT-SFT \citep{BAT2025} & Binaural & Semantic judge & 26.23 & 21.51 & 27.08 & 25.20 \\
      Qwen2-Audio-7B \citep{qwen2audio2024} & Mono & Semantic judge & 52.40 & 42.40 & 48.80 & 48.40 \\
      Qwen2.5-Omni-7B \citep{qwen25omni2025} & Mono & Exact avg. & 58.44 & 49.64 & 66.97 & 58.85 \\
      Step-Audio-R1.1 & Mono & Exact parse & 59.22 & 48.03 & 71.13 & 60.10 \\
      Gemini-3.1-Pro Preview & Mono & Exact parse & 57.92 & 54.12 & 74.11 & 62.30 \\
      \midrule
      TWNM-SAPO & FOA & Exact MCQA & \textbf{63.64} & \textbf{69.89} & \textbf{79.76} & \textbf{70.80} \\
      \bottomrule
    \end{tabularx}%
\end{table}

\blue{The matched controls distinguish text priors, monaural acoustic evidence, and the contribution of the spatial pathway. Question-only decoding scores 28.2\%. A trained $W$-only model uses the same frozen decoder base, Stage~2/3 examples, optimization schedule, batch size, update count, decoding, and scorer as Full SFT, while removing the spatial encoder, projection, and normalization throughout training. The $W$ channel retains event, speech, loudness, reverberation, and count cues, and the trained model scores 49.3\%. Adding the full FOA spatial pathway raises accuracy to 62.0\%, a paired gain of 12.7 percentage points (95\% CI $[9.2,16.1]$).}

\Needspace{5\baselineskip}
\blue{The gain measures the additional contribution of the spatial pathway after the monaural system has adapted to its available evidence. SAPO reaches 70.8\% (paired $+8.8$ points, 95\% CI $[5.6,12.1]$), improving the decoder's effective use of fixed spatial-semantic evidence while both encoders and the projector remain frozen (Appendix~\ref{app:matched_controls}).}

\blue{Monaural LALMs receiving only the omnidirectional channel $W$ use semantic and non-directional acoustic evidence for event, speech, loudness, and reverberation questions. Their performance decreases on families requiring source-location binding and listener-centric geometric reasoning. BAT uses binaural audio and single-source spatial-QA supervision; TWNM processes full FOA for multi-source ASA. Table~\ref{tab:main_results_transposed} reports results under the specified input and answer interfaces, with complete audits in Appendix~\ref{app:additional_baselines}.}

\blue{Table~\ref{tab:spatial_semantic_split} separates spatial and semantic-heavy families. Monaural models score higher on semantic-heavy questions, where event identity, speech, and other non-directional acoustic cues support the answer. Source-location binding, observer rotation, and relative-direction judgments require source-specific geometry carried by FOA's directional channels. TWNM reaches 66.4\% on spatial families while remaining competitive on semantic-heavy questions, showing the complementary roles of semantic content and spatial structure in ASA.}

\begin{table}[t]
  \centering
  \caption{Spatial versus semantic decomposition on the ASA benchmark. Rows use the audit protocol associated with the same model in Table~\ref{tab:main_results_transposed} or Appendix~\ref{app:additional_baselines}, and values are accuracy percentages.}
  \label{tab:spatial_semantic_split}
  {\small
    \begin{tabular}{@{}lcc@{}}
      \toprule
      \textbf{Model} & \textbf{Spatial families (\%)} & \textbf{Semantic families (\%)} \\
      \midrule
      Qwen2-Audio-7B & 42.63 & 55.63 \\
      Step-Audio-R1.1 & 45.32 & \textbf{78.60} \\
      Qwen3-Omni-Flash & 46.58 & 71.85 \\
      Qwen2.5-Omni-7B & 47.30 & 73.31 \\
      GPT-audio & 50.36 & 68.69 \\
      Gemini-3.1-Pro Preview & 49.28 & \textbf{78.60} \\
      TWNM-SAPO & \textbf{66.37} & 76.35 \\
      \bottomrule
    \end{tabular}%
  }
\end{table}

\Needspace{5\baselineskip}
\blue{Because each family contains only 55--56 items, the per-family breakdowns in Table~\ref{tab:main_task_family_results} are exploratory diagnostics; the primary quantitative conclusions rest on the full 1,000-item set and the $\mathcal{L}_1/\mathcal{L}_2/\mathcal{L}_3$ aggregates. Input-perturbation controls in Appendix~\ref{app:control_perturbations} distinguish question and option priors from the acoustic evidence available to the model.}

\begin{table}[t]
  \centering
  \caption{Task-family accuracy on the ASA benchmark. Values are percentages; BAT-SFT is a diagnostic reference using semantic-answer audit, while TWNM stages use exact final-option MCQA matching. Short labels are expanded in Appendix~\ref{app:prompt}.}
  \label{tab:main_task_family_results}
  {\small
      \begin{tabular}{@{}clrrrr@{}}
        \toprule
        \textbf{Level} & \textbf{Task family} & \textbf{BAT-SFT} & \textbf{TWNM-Align} & \textbf{TWNM-SFT} & \textbf{TWNM-SAPO} \\
        \midrule
        \multirow{7}{*}{$\mathcal{L}_1$}
        & Count & 36.36 & 65.45 & \textbf{67.27} & 60.00 \\
        & Acoustics & 7.27 & 61.82 & \textbf{78.18} & 69.09 \\
        & Distance & 25.45 & 38.18 & 56.36 & \textbf{70.91} \\
        & Event class & 18.18 & 61.82 & 58.18 & \textbf{83.64} \\
        & Speech & 21.82 & 41.82 & \textbf{80.00} & 60.00 \\
        & Azimuth & 27.27 & \textbf{54.55} & 52.73 & \textbf{54.55} \\
        & Elevation & 47.27 & 49.09 & \textbf{63.64} & 47.27 \\
        \midrule
        \multirow{5}{*}{$\mathcal{L}_2$}
        & Binding & 12.50 & 16.07 & \textbf{85.71} & 71.43 \\
        & Distance comparison & 25.00 & 35.71 & 64.29 & \textbf{82.14} \\
        & ID from location & 25.00 & 42.86 & 66.07 & \textbf{91.07} \\
        & Location from ID & 28.57 & 16.07 & \textbf{80.36} & 58.93 \\
        & Relative direction & 16.36 & 36.36 & \textbf{56.36} & 45.45 \\
        \midrule
        \multirow{6}{*}{$\mathcal{L}_3$}
        & Causal intent & 19.64 & 51.79 & 41.07 & \textbf{82.14} \\
        & Source removal & 44.64 & 32.14 & 53.57 & \textbf{78.57} \\
        & Rotate $180^\circ$ & 42.86 & 46.43 & 67.86 & \textbf{69.64} \\
        & Multi-hop & 10.71 & 26.79 & 30.36 & \textbf{71.43} \\
        & Physical consistency & 25.00 & 37.50 & 76.79 & \textbf{89.29} \\
        & Scene abduction & 19.64 & 64.29 & 37.50 & \textbf{87.50} \\
        \bottomrule
      \end{tabular}%
  }
\end{table}

\blue{The $\mathcal{L}_2$ families test whether identity and location belong to the same source. Recognizing an event and estimating a direction elsewhere in the mixture leaves this correspondence unresolved. SFT raises identity--location--distance binding from 16.07\% to 85.71\% and location retrieval from identity from 16.07\% to 80.36\%. SAPO further improves distance comparison and identity retrieval from location, while some binding and directional families decrease. At $\mathcal{L}_3$, gains are pronounced in scene abduction, physical consistency, causal intent, and multi-hop composition. These task patterns distinguish the benefits of supervised answer alignment from the later improvement in scene-level reasoning.}

\FloatBarrier
\Needspace{8\baselineskip}
\subsection{Training Diagnostics}
\label{sec:ablation_summary}

\begin{table}[H]
  \centering
  \caption{TWNM training-stage diagnostics under exact MCQA scoring. Values are percentages.}
  \label{tab:main_diagnostics}
  {\small
    \begin{tabular}{@{}lcccc@{}}
      \toprule
      \textbf{Stage} & $\mathcal{L}_1$ & $\mathcal{L}_2$ & $\mathcal{L}_3$ & \textbf{All} \\
      \midrule
      TWNM-Align & 53.25 & 29.39 & 43.15 & 43.20 \\
      TWNM-SFT & \textbf{65.19} & \textbf{70.61} & 51.19 & 62.00 \\
      TWNM-SAPO & 63.64 & 69.89 & \textbf{79.76} & \textbf{70.80} \\
      \bottomrule
    \end{tabular}
  }
\end{table}

\blue{Table~\ref{tab:main_diagnostics} reports the submitted training-stage trajectory. Supervised fine-tuning produces the largest single-level gain at $\mathcal{L}_2$, raising relational-integration accuracy from 29.39\% after alignment to 70.61\%, and increasing the overall score from 43.20\% to 62.0\%. This stage learns to associate the spatial and semantic tokens produced by the frozen encoders with ASA question--answer pairs. SAPO updates only decoder LoRA while both encoders and the projector remain fixed. It raises overall accuracy from 62.0\% to 70.8\% and $\mathcal{L}_3$ scene reasoning from 51.19\% to 79.76\%. In this stage trajectory, $\mathcal{L}_1$ decreases from 65.19\% to 63.64\% and $\mathcal{L}_2$ from 70.61\% to 69.89\%. The stages therefore have distinct effects across the hierarchy: SFT supplies the main relational-integration gain, while SAPO improves end-to-end use of the fixed evidence, with its largest gain in scene reasoning. Detailed projector and reward diagnostics appear in Appendix~\ref{app:ablation_details}.}

\subsection{Real-Recording Diagnostics}
\label{sec:starss23_qa}

\blue{STARSS23 provides a corpus-level real-recording diagnostic following Stage~1 sim-to-real adaptation. TWNM-SAPO reaches 44.0\% overall on a QA set excluded from LALM SFT and SAPO. Replacing the FOA waveform with Gaussian noise under the same questions reduces accuracy to 25.8\%, demonstrating end-to-end acoustic sensitivity. Together with physical-variable recovery in Section~\ref{sec:encoder_analysis}, these results show that the learned audio interface uses real-recording evidence (Table~\ref{tab:starss23_qa}; Appendix~\ref{app:starss_details}).}

\begin{table}[H]
  \centering
  \caption{End-to-end STARSS23 QA diagnostic. Values are generated-answer audit accuracies over attempted questions.}
  \label{tab:starss23_qa}
  {\small
    \begin{tabular}{@{}lccccc@{}}
      \toprule
      \textbf{Run} & \textbf{$N$} & $\mathcal{L}_1$ (\%) & $\mathcal{L}_2$ (\%) & $\mathcal{L}_3$ (\%) & \textbf{Overall (\%)} \\
      \midrule
      Qwen2.5-Omni-7B & 500 & 41.90 & \textbf{40.48} & 36.89 & 40.20 \\
      TWNM-SFT & 500 & 41.90 & 38.10 & 47.54 & 42.00 \\
      TWNM-SAPO & 500 & \textbf{45.24} & 38.69 & \textbf{49.18} & \textbf{44.00} \\
      TWNM-SAPO, Gaussian noise & 500 & 21.90 & 25.60 & 32.79 & 25.80 \\
      \bottomrule
    \end{tabular}%
  }
\end{table}

\blue{The encoder and full-policy evaluations test complementary uses of real audio. Source count, direction, and distance measure physical-variable recovery at the acoustic interface. The QA contrast between real FOA and Gaussian noise tests whether the complete policy remains sensitive to the recording while the question stays fixed. The 500 QA items are excluded from LALM SFT and SAPO, while STARSS23 also supplies Stage~1 adaptation data. The results therefore establish physical recovery and end-to-end acoustic sensitivity within this corpus. Appendix~\ref{app:starss_details} gives the question construction and scoring protocol.}

\section{Conclusion}
\label{sec:conclusion}

We formulated spatial ASA as a three-level problem spanning atomic perception, relational integration, and cognitive reasoning. TWNM combines FOA-conditioned spatial encoding, dense spatial-semantic projection, and progressive alignment, while the ASA benchmark provides auditable metadata-grounded evaluation beyond source recognition. The main lesson is that spatial audio-language reasoning should be evaluated as binding and composition, not only localization: monaural LALMs remain strong on semantic-heavy families, but weaken when answers require directional channels, source-wise assignment, listener-centric relations, or counterfactual viewpoint changes. Stage-3 tuning gives the main binding gain, \blue{SAPO improves the decoder's effective use of fixed spatial-semantic evidence}, and corrupted-input controls show measurable priors but a need for intact FOA spatial evidence.

\paragraph{Limitations.}
The synthetic ASA benchmark provides auditable source identity, location, distance, room, and counterfactual labels, but its holdout split does not separate dry-clip identities because the LALM render corpora share public source datasets. STARSS23 \blue{provides a corpus-level diagnostic}, while longer recordings, moving sources or listeners, denser scenes, and in-the-wild spatial QA remain open. \blue{The encoder has a fixed capacity of $K=3$ sources.} TWNM assumes multichannel spatial audio, and releases should preserve provenance and privacy-use restrictions.

\paragraph{Reproducibility.}
Appendix~\ref{app:training_details} reports settings; the supplement provides a compact architecture prototype, while full corpora, checkpoints, closed-API audits, and cluster-specific scripts are not bundled.

\begin{ack}
\blue{This work was supported by the National Key Research and Development Program of China (No. 2024YFB2808902) and the High-Performance Computing Platform of Peking University. The authors declare no competing interests.}
\end{ack}

\newpage
\bibliographystyle{plainnat}
\bibliography{example_paper,related-work-additions}

\newpage
\appendix
\raggedbottom
\pretocmd{\section}{\FloatBarrier}{}{}
\pretocmd{\subsection}{\Needspace{6\baselineskip}}{}{}
\setlength{\textfloatsep}{12pt plus 2pt minus 2pt}
\setlength{\intextsep}{10pt plus 2pt minus 2pt}
\section*{Appendix}

\section{Limitations and Societal Impact}
\label{app:limitations}

The study focuses on controlled FOA simulation, Stage-1 real-world adaptation on STARSS23, and an end-to-end STARSS23 QA set. The synthetic ASA benchmark is the primary evaluation because simulator metadata gives precise source identities, coordinates, distances, and room acoustics. The real-recording QA set covers a fixed subset of STARSS23-derived questions, while long-duration and unconstrained in-the-wild spatial QA require broader annotation and evaluation. The ASA controls in Appendix~\ref{app:control_perturbations} show that language and option priors remain measurable even when the audio is silenced, mismatched, or stripped to the $W$ channel, so the reported gains should be interpreted within the simulator-policy setting rather than as unrestricted real-world spatial competence. The method assumes access to multichannel spatial audio, so direct deployment on monaural consumer recordings would require additional capture or spatial reconstruction.

The work may benefit robotics, augmented reality, assistive listening, and spatial scene understanding. At the same time, stronger audio localization and spatial reasoning could be misused for acoustic surveillance or privacy-invasive inference. Any public release of models, generated datasets, or benchmark assets should therefore include clear documentation of intended research use, dataset provenance, and restrictions against privacy-invasive deployment.

\section{LLM Usage Statement}
\label{app:llm_usage}

LLMs are part of the core research object and method: the proposed system uses an LLM decoder for audio-language reasoning, and a teacher LLM is used during benchmark construction to turn simulator-derived RTSD metadata into multiple-choice QA items. The simulator metadata provides the ground-truth spatial facts and answer labels. The teacher LLM is not used to infer hidden labels from audio. Main TWNM scores on the synthetic ASA benchmark are computed by exact MCQA matching; LLM judges are used for free-form baseline audits and STARSS23 QA where exact option extraction is not sufficient.

\section{Formal ASA Task Definition}
\label{app:formal_asa}

An auditory scene $\mathcal{S}$ is a composition of \blue{auditory objects} and scene-level acoustic attributes observed through a multichannel waveform $\mathbf{X} \in \mathbb{R}^{C \times L}$. Let $\mathcal{O}=\{o_1,\ldots,o_N\}$ denote the set of \blue{source and background/environment objects, as in Section~\ref{sec:problem_formulation}}. Each object is represented as $o_i=(c_i,\mathbf{s}_i,\tau_i)$, where $c_i$ is a semantic class or transcript-like attribute, $\mathbf{s}_i$ is a spatial coordinate such as azimuth, elevation, and distance relative to the listener, \blue{or a null coordinate for diffuse environment objects,} and $\tau_i$ is the temporal support of the event. \blue{Reverberation and room conditions are scene-level attributes.}

At $\mathcal{L}_1$, the target is a set of candidate attributes $\hat{\mathcal{A}}=f_{\mathcal{L}_1}(\mathbf{X})$ containing semantic, spatial, temporal, and room-acoustic evidence. At $\mathcal{L}_2$, the target is a relational scene abstraction $\mathcal{G}=(\mathcal{V},\mathcal{E})$, where nodes correspond to validated objects and edges encode relations such as relative direction, distance comparison, co-occurrence, or physical consistency. Attribute binding can be written as $\hat{\bar{\mathbf{a}}}=\operatorname*{argmax}_{\bar{\mathbf{a}}}P(\bar{\mathbf{a}}\mid\mathbf{a},\mathbf{X})$, where the model infers missing spatial, semantic, or temporal attributes from observed evidence. At $\mathcal{L}_3$, the target answer is a query-conditioned judgment $Y^*=\operatorname*{argmax}_Y P(Y\mid q,\mathcal{G},\mathcal{K},\mathbf{X})$, where $\mathcal{K}$ denotes implicit physical and scene priors encoded in the language model rather than an external rule base.

\section{Data and Benchmark Construction}
\label{app:data_pipeline}
\label{app:data_details}

\subsection{Pipeline, Corpora, and Holdout Definition}

\begin{figure}[htbp]
  \centering
  \includegraphics[width=\linewidth]{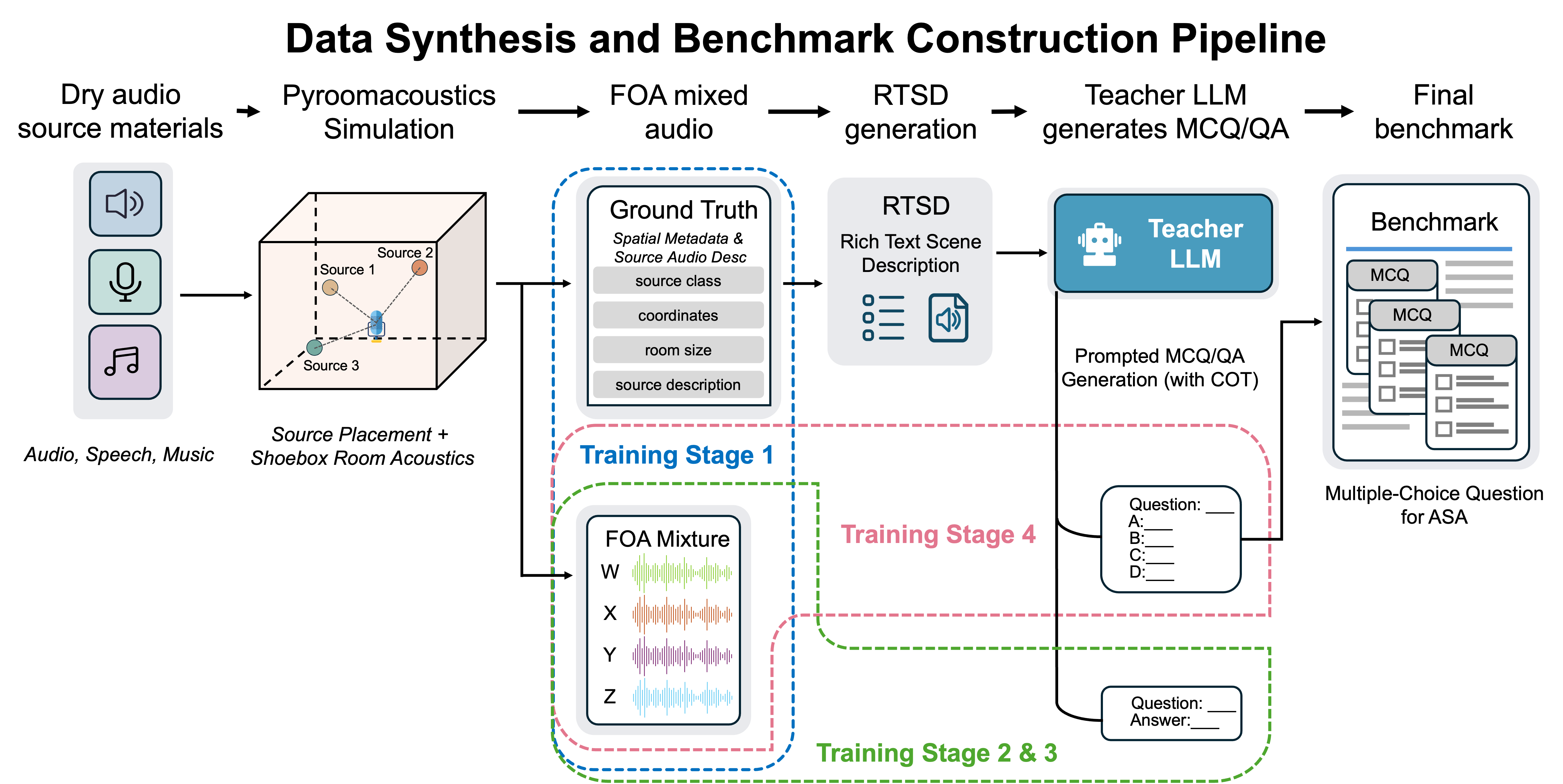}
  \caption{Data synthesis, training-data generation, and benchmark construction pipeline. Dry source clips are spatialized with Pyroomacoustics, converted into FOA mixtures with simulator-derived metadata, serialized into RTSDs, and transformed into training or benchmark QA items.}
  \label{fig:data_pipeline}
\end{figure}

Figure~\ref{fig:data_pipeline} summarizes the full data path used by TWNM. Dry source clips are first placed into physically controlled shoebox rooms, rendered as spatial audio, and saved together with simulator-derived metadata. The metadata is the single source of ground truth for spatial supervision: it records source identities, room conditions, source coordinates in the room, relative vectors from the listener, azimuth, elevation, distance, and the output audio paths. These facts are then serialized into rich text scene descriptions (RTSDs), which are used to build training prompts and held-out benchmark questions. During evaluation, the model receives only the audio and the question; the RTSD, simulator metadata, and answer key are never exposed to the model.

The data pipeline uses separate synthetic corpora with the same physical scene sampler but different experimental roles. The 50k full-simulation corpus is used for controlled spatial encoder experiments and isolates event-level spatial perception with FSD50K supervision. The SFT, SAPO, and benchmark render corpora use a broader event/speech/music source pool and are rendered into separate scene directories. The benchmark corpus is held out at the rendered-scene, spatial-configuration, RTSD, question, and answer-key level from the LALM training batches, while dry-source clip identities can overlap because these corpora are drawn from the same public source datasets. Thus, the benchmark tests held-out spatial compositions, room draws, rendered mixtures, questions, and answer keys rather than unseen dry-clip identity.

The archived metadata confirms this holdout scope. Among the 10,154 unique dry clips appearing in the benchmark render corpus, 5,911 also appear in the SFT render metadata and 5,967 also appear in the SAPO render metadata. The benchmark split is therefore defined over rendered mixtures, geometry, room draws, RTSDs, questions, and answers, not over dry-source identity.

The final 1,000-question ASA test set is selected from the 10k benchmark render corpus and excluded from all SFT and SAPO batches. Evaluation metadata stores a unique benchmark-corpus sample ID, rendered FOA path, task family, level, RTSD, MCQ instruction, and gold answer. Options are randomized during generation, malformed or ambiguous LLM generations are rejected, and correctness is tied back to simulator metadata rather than to the teacher model's preference. The benchmark package includes the metadata needed to audit the scene IDs, questions, and answer keys used for final reporting.

\begin{table}[htbp]
  \centering
  \blue{\caption{Synthetic corpora used in the data pipeline. The render corpora use the same room, source-placement, and spatial-rendering policy, but they serve different experimental roles.}
  \label{tab:synthetic_corpora}}
  \small
  \begin{tabularx}{\linewidth}{@{}p{0.24\linewidth}X@{}}
    \toprule
    \textbf{Corpus} & \textbf{Role and construction} \\
    \midrule
    Full-simulation corpus & 50,000 scenes and 75,150 source instances for controlled spatial encoder training and analysis. Dry events are sampled from FSD50K development audio \citep{fsd50k}; labels are mapped to the official 200-class vocabulary, with multi-label event annotations preserved. \\
    SFT render corpus & 50,000 scenes and 84,781 source instances for supervised audio-language training, producing 45,000 instruction-tuning examples after QA filtering. Source domains are sampled with event/speech/music weights of 5:3:2 from ClothoV2, LibriTTS-R, and SongDescriber. \\
    SAPO render corpus & 50,000 scenes and 84,866 source instances for preference-optimization rollouts. It uses the same public source pools and simulator policy as SFT, but separate rendered scenes and RTSDs. \\
    Benchmark render corpus & 10,000 scenes and 16,744 source instances for held-out ASA benchmark construction. The final 1,000-question test set is selected from this corpus and is excluded from all SFT and SAPO batches. \\
    \bottomrule
  \end{tabularx}
\end{table}

\subsection{Simulation and Rendering Details}

\paragraph{Room and source sampling.}
Room size, reverberation, and source count are sampled independently. Small, medium, and large rooms use length and width ranges of $[3,5]$ m, $[5,8]$ m, and $[8,15]$ m, respectively, with sampling probabilities of 0.6, 0.3, and 0.1. Height is sampled independently from $[2.5,4.0]$ m. In the event/speech/music render corpora, each scene contains one, two, or three sources with probabilities 0.5, 0.3, and 0.2. The full-simulation corpus uses the same spatial sampler but reports the retained source-instance count after FSD50K clip filtering and successful rendering. The listener is placed at $(L/2,W/2,\min(1.7,H-0.5))$, and each source is sampled uniformly inside the room with a 0.5 m boundary margin on each axis. For the event/speech/music render corpora, each source is additionally assigned a mixing gain sampled from $[-6,3]$ dB before summation.

Because source gain is randomized, metric distance in the synthetic QA tasks evaluates metadata-controlled distance reasoning under the fixed simulator policy. Absolute ranging from audio alone can be confounded by source loudness and direct-to-reverberant structure, so distance questions test the model's use of simulated spatial evidence and metadata-derived labels.

Reverberation is controlled through a room-level energy absorption coefficient $\alpha$. The three reverberation regimes are weak, mid, and strong, with $\alpha \in [0.7,0.999]$, $[0.3,0.7]$, and $[0.1,0.3]$, respectively. For metadata, the estimated reverberation time follows the Sabine formula
\begin{equation}
  \widehat{T}_{60} = \frac{0.161 V}{S \alpha},
\end{equation}
where $V=LWH$ is room volume and $S=2(LW+LH+WH)$ is surface area. The implementation clips $\alpha$ to $[0.01,0.99]$ for the RT60 calculation. The simulator stores both Cartesian and spherical source coordinates, so localization, distance estimation, spatial comparison, and counterfactual observer-rotation questions can all be derived from the same scene state rather than from separate annotations.

\paragraph{Spatial rendering.}
The simulator renders each scene with \texttt{pyroomacoustics} \citep{pyroomacoustics2017} using a shoebox room and image-source order 15 \citep{allen1979image}, with ray tracing disabled, following the broader use of geometric acoustic simulation in spatial and embodied audio environments \citep{soundspaces2020}. The FOA renderer uses one omnidirectional component and three co-located first-order dipole components along the lateral, vertical, and front-back axes. We use azimuth $0^\circ$ as front, $90^\circ$ as right, $180^\circ$ as back, and $270^\circ$ as left; elevation is positive upward. The stored channel order is ACN-style $[W,Y,Z,X]$, and the simulator, metadata labels, and model inputs use the same fixed sign convention throughout \citep{aes69sofa2015,rafaely2015fundamentals,you2025}. A matched binaural rendering is also saved for compatible baseline analyses, but the TWNM experiments in this paper use the FOA stream as the spatial audio input.

\subsection{RTSD and Held-Out Benchmark Generation}

\paragraph{RTSD and benchmark construction.}
Each scene metadata file is converted into an RTSD with deterministic formatting. Numeric spatial values are rounded to two decimals. Azimuth is mapped to eight named sectors using 22.5-degree offsets: front, front-left, left, back-left, back, back-right, right, and front-right. Elevation is converted to above, level, or below using a $\pm 10^\circ$ threshold. Distance is converted to near, mid, or far using thresholds of 1.2 m and 2.0 m. The RTSD also includes room size, reverberation type, estimated RT60, absorption, and the complete list of active sources. This representation provides auditable evidence for QA generation while keeping evaluation audio-only.

The held-out ASA benchmark contains 1,000 multiple-choice questions built from the benchmark render corpus. The benchmark is stratified into 385 $\mathcal{L}_1$ questions, 279 $\mathcal{L}_2$ questions, and 336 $\mathcal{L}_3$ questions, covering the task families listed in Table~\ref{tab:benchmark_tasks}. The teacher LLM is used to phrase questions and options from the RTSD, while correctness remains tied to simulator metadata. The archived May 2026 generation run uses Gemini 3 Flash through an OpenAI-compatible API with retries and strict JSON parsing. The generation pipeline requires four options, a single answer letter, audio-only wording that does not mention the RTSD, and rejection of scenes that cannot support an unambiguous question. This design keeps the natural-language surface diverse without making the teacher model the source of spatial labels.

\section{Training and Reproducibility Details}
\label{app:training_details}

\subsection{Stage 1 Spatial Encoder}
\label{app:spatial_encoder_results}
The spatial encoder is trained before the LALM alignment stages. Table~\ref{tab:stage1_hyperparams} summarizes the hyperparameters used for the encoder analysis in Section~\ref{sec:encoder_analysis}. Phase I trains on the full-simulation corpus with complete synthetic labels. Phase II initializes from the Phase I checkpoint and mixes simulation with STARSS23 real FOA clips at a 7:3 batch ratio; semantic-class, room-area, and absorption losses are masked for real clips because STARSS23 does not provide the same vocabulary and room-acoustic labels as the simulator. The real clips used for Phase II come from the STARSS23 development-training splits, while the encoder metrics in Table~\ref{tab:encoder_diagnostics} use development-evaluation splits.

\begin{table}[htbp]
  \centering
  \blue{\caption{Stage 1 spatial-encoder training details. Both phases use an 8-layer encoder with hidden dimension 96 and at most three source slots.}
  \label{tab:stage1_hyperparams}}
  \small
  \begin{tabularx}{\linewidth}{@{}lXX@{}}
    \toprule
    \textbf{Setting} & \textbf{Phase I: synthetic warm-up} & \textbf{Phase II: sim-to-real adaptation} \\
    \midrule
    Data & 50k full-simulation corpus & Simulation plus STARSS23 FOA clips \\
    Sample rate / clip length & 44.1 kHz / 1.0 s & 44.1 kHz / 1.0 s \\
    Max sources & 3 & 3 \\
    Optimizer / LR & AdamW / $1\times10^{-4}$ & AdamW / $3\times10^{-5}$ \\
    Losses & Direction, distance, class, activity, count, room area, absorption & Direction, distance, activity, and count on real clips; full losses on synthetic clips \\
    \bottomrule
  \end{tabularx}
\end{table}

Table~\ref{tab:encoder_diagnostics} reports the Stage-1 spatial encoder results used in the main paper. For STARSS23 \citep{starss23}, Stage-1 sim-to-real adaptation uses development-training recordings, while evaluation uses static one-second FOA segments extracted from official development-evaluation splits. A segment is retained only when the active source set, azimuth, elevation, distance, and class labels remain constant over the corresponding 100 ms annotation frames; empty frames and clips with more than three active sources are excluded. Predicted and ground-truth slots are matched with Hungarian assignment using direction, distance, and objectness costs.

\subsection{LALM Training Hyperparameters}
We use AdamW for the LALM alignment stages and LoRA adapters for decoder tuning \citep{hu2022lora}. Table~\ref{tab:hyperparams} summarizes the optimization settings that define the reported runs. Stage 2 aligns only the dense hybrid projector, Stage 3 tunes the projector together with decoder LoRA adapters, and Stage 4 applies SAPO post-training on top of the supervised model.

\begin{table}[htbp]
  \centering
  \blue{\caption{Optimization hyperparameters for the LALM training stages.}
  \label{tab:hyperparams}}
    \begin{tabularx}{\columnwidth}{@{}lYYY@{}}
      \toprule
      \textbf{Hyperparameter} & \textbf{Stage 2} & \textbf{Stage 3} & \textbf{Stage 4} \\
      \midrule
      Trainable modules & Dense hybrid projector & Projector and decoder LoRA & SAPO LoRA adapters \\
      Learning rate & $5 \times 10^{-5}$ & $5 \times 10^{-5}$ & $1 \times 10^{-6}$ \\
      LoRA Rank ($r$) & N/A & 8 & 8 \\
      LoRA Alpha ($\alpha$) & N/A & 32 & 32 \\
      \bottomrule
    \end{tabularx}
\end{table}

\subsection{SAPO Reward and Optimization}
For each query, we generate $G=8$ rollouts and compute advantages using a fine-grained, weighted reward function:
\begin{equation}
  \label{eq:reward}
  R(y) = 2.0 r_{\text{fmt}} + 1.0 r_{\text{val}} + 3.0 r_{\text{cor}} + 0.5 r_{\text{len}} + 0.5 r_{\text{evd}} + 0.2 r_{\text{ref}} .
\end{equation}
The reward combines six components: $r_{\text{fmt}}$ enforces the XML-style reasoning structure, $r_{\text{val}}$ validates that the output is a legitimate option (A--D), $r_{\text{cor}}$ rewards prediction accuracy against the ground truth, $r_{\text{len}}$ penalizes reasoning chains outside the 40--400 character target window, $r_{\text{evd}}$ is a lightweight evidence-shaping term that encourages the answer to refer to audible evidence, and $r_{\text{ref}}$ is a non-positive penalty for referencing the hidden text prompt or RTSD rather than the audio. The correctness reward is the dominant term; the evidence and reference terms are auxiliary safeguards against fluent but ungrounded rationales.

To reduce reward hacking risk and keep Kullback--Leibler (KL) drift bounded during optimization, we implement an adaptive KL penalty. The SAPO run initializes $\beta=0.02$, targets a moving-average KL of 1.0, clamps $\beta$ to $[0.0,0.2]$, and adjusts it with multiplicative scale 1.5. We use token-level importance sampling, rollout temperature 1.0, top-$p=0.95$, SAPO temperatures $\tau_{+}=1.0$ and $\tau_{-}=1.05$, and a 256-token generation cap. The Stage 4 LoRA adapters are initialized with a Gaussian distribution rather than zero.

\paragraph{Implementation notes.}
TWNM is implemented in PyTorch and Hugging Face Transformers \citep{paszke2019pytorch,wolf2020transformers}. The reported model training used distributed GPU execution with 32 GB-class accelerators and DeepSpeed ZeRO-style memory partitioning \citep{rajbhandari2020zero}. The anonymized supplement provides a compact architecture prototype containing the core TWNM model code, spatial-encoder wrapper, dense hybrid projector, Audio-Flamingo-3 component loader, and representative architecture configuration. It is intended to support inspection and prototype-level reuse of the architecture; paper-scale training, evaluation audits, closed-API baselines, full corpus rendering, and large checkpoints require external raw assets, provider access, and cluster-specific storage not bundled with the submission.

\section{Evaluation Protocols and Complete ASA Results}
\label{app:evaluation_protocols}

This appendix collects the benchmark taxonomy, prompt contracts, automatic scoring rules, input-perturbation controls, complete ASA results, external-baseline audit matrix, and BAT-specific diagnostics. The goal is to keep the evaluation evidence chain in one place: first defining what is evaluated, then how outputs are scored, and finally reporting the complete numbers behind the compact main tables.

Benchmark construction uses a shared MCQ output contract. The generator must use English, refer only to the audio, avoid mentioning RTSD or scene descriptions, produce exactly four options, shuffle the correct option, and output strict JSON with a single gold answer letter. If the RTSD is insufficient for an unambiguous question, the generator returns \texttt{UNABLE\_TO\_GENERATE}. A separate teacher pass then expands the gold letter into the reasoning trace used for SFT.

\subsection{Prompt Templates and Benchmark Examples}
\label{app:prompt}

The following prompt skeleton is the template family used for benchmark MCQ generation; task-specific files add the skill definition and distractor rules for each family in Table~\ref{tab:benchmark_tasks}.

\begin{promptbox}[Shared MCQ generation prompt skeleton]
  {\small\setlength{\parindent}{0pt}\raggedright
    Task: based on the Scene Description, create exactly one audio-only four-choice MCQ sample.\\[2pt]
    Rules: use English only; do not mention any text, description, or RTSD; provide exactly four shuffled options (A--D); output strict JSON only; if the scene is ambiguous or insufficient, output \texttt{UNABLE\_TO\_GENERATE}.\\[2pt]
    Skill: \texttt{\{task-specific skill, e.g., bind identity to location and distance\}}.\\
    Requirements: \texttt{\{task-specific target, option, and distractor rules\}}.\\[2pt]
    JSON: \texttt{\{"instruction": "<question>\textbackslash nA) ...\textbackslash nB) ...\textbackslash nC) ...\textbackslash nD) ...", "answer": "X"\}}\\[2pt]
    Scene Description: \texttt{\{\{ rtsd \}\}}
  }
\end{promptbox}

\begin{promptbox}[Teacher reasoning prompt for MCQ SFT]
  {\small\setlength{\parindent}{0pt}\raggedright
    You are a teacher model that writes a short reasoning trace for a given MCQ and its correct answer. The input includes a Scene Description for internal reasoning only, the question, options A--D, and the correct answer. Output must be a single line in this exact format:\\[2pt]
    \texttt{|<think>| ... |</think>| ... |<answer>| X |</answer>|}\\[2pt]
    The think block must restate the question focus, cite audio evidence that supports the correct option, explain why the other options are incorrect, and stay under 200 words. The final answer block must contain only the correct option letter.
  }
\end{promptbox}

\begin{figure}[!htbp]
  \centering
  \includegraphics[width=\linewidth]{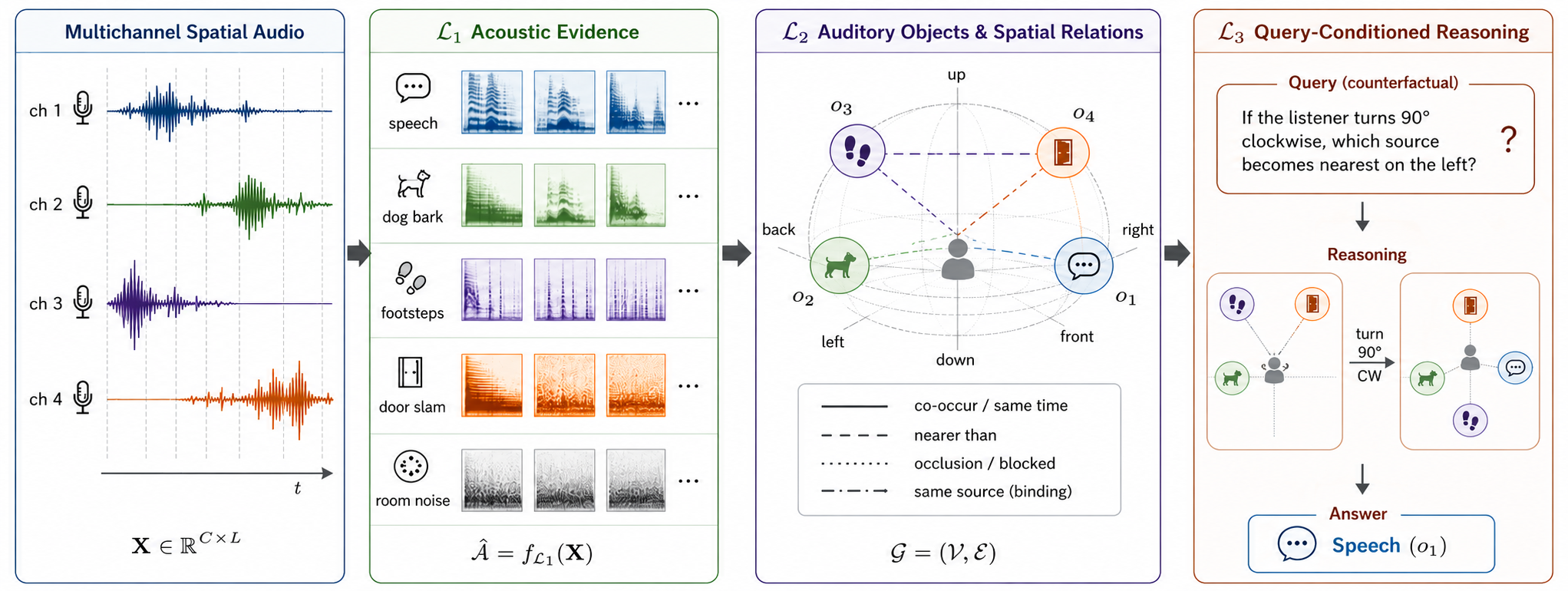}
  \caption{Schematic examples from the ASA benchmark. The panels illustrate how the same benchmark protocol spans atomic perception, relational integration, and cognitive reasoning. In each example, A--D denote the four answer choices and the highlighted choice is the illustrative gold option.}
  \label{fig:benchmark_examples}
\end{figure}

Figure~\ref{fig:benchmark_examples} gives reader-facing examples for the three ASA levels. These schematics are not model inputs; evaluation remains audio-only with text questions and options, while hidden simulator metadata determines the gold answer.

\FloatBarrier
\subsection{Automatic Audit Protocols}
\label{app:audit_protocols}

For exact-option audits, the parser operates on the generated response text after removing chat-template terminators. It first searches for the tagged answer format used by TWNM training, then for explicit phrases such as ``answer'', ``option'', ``choice'', or ``final answer'' followed by A--D. If no valid option letter can be extracted, the sample is counted as incorrect without manual repair. For semantic-answer audits, the judge is not asked to solve the audio task; it only checks whether a model's free-form answer semantically matches the gold option text for the same MCQ.

\blue{Under a binomial sampling model, the two-sided 95\% Wilson interval for accuracy $\hat p=x/n$ has half-width $h=\frac{z}{1+z^2/n}\sqrt{\hat p(1-\hat p)/n+z^2/(4n^2)}$, with $z=1.96$. For TWNM's $\hat p=708/1000$ on the fixed ASA evaluation, this gives approximately 2.8 percentage points.}

\begin{promptbox}[Semantic-answer audit prompt skeleton]
  {\small\setlength{\parindent}{0pt}\raggedright
    You are auditing an answer to a four-choice audio question. You do not have access to the audio and should not infer a new answer. Decide only whether the Model Answer semantically matches the Gold Option.\\[2pt]
    Question: \texttt{\{\{ question \}\}}\\
    Options: \texttt{\{\{ options \}\}}\\
    Gold Option: \texttt{\{\{ gold\_letter \}\}. \{\{ gold\_text \}\}}\\
    Model Answer: \texttt{\{\{ model\_answer \}\}}\\[2pt]
    Mark correct if the model clearly selects or paraphrases the gold option, even without the option letter. Mark incorrect if it selects another option, gives multiple incompatible choices, refuses, is ambiguous, or lacks enough information. Output strict JSON only: \texttt{\{"correct": true/false\}}.
  }
\end{promptbox}

\FloatBarrier
\subsection{Benchmark Taxonomy}
\label{app:benchmark_taxonomy}

\blue{The spatial families contain 556 questions covering distance estimation, azimuth/elevation localization, identity--location--distance binding, distance comparison, identity--location retrieval in both directions, relative direction, observer rotation, and multi-hop spatial composition. The semantic-heavy families contain 444 questions covering object counting, environment acoustics, event recognition, speech transcription, causal intent, source removal, physical consistency, and scene abduction.}

{\small
\setlength{\LTcapwidth}{\linewidth}
\begin{longtable}{@{}p{0.09\linewidth}p{0.17\linewidth}p{0.23\linewidth}p{\dimexpr0.51\linewidth-6\tabcolsep\relax}@{}}
  \caption{Overview of the TWNM spatial reasoning benchmark. The benchmark covers three cognitive levels ($\mathcal{L}_1$--$\mathcal{L}_3$), derived from the ASA framework. Task-specific prompts use simulator metadata to create audio-only MCQs with a single gold answer.}
  \label{tab:benchmark_tasks}\\
      \toprule
      \textbf{Level} & \textbf{Task family} & \textbf{Specific tasks} & \textbf{Example question template} \\
      \midrule
      \endfirsthead
      \toprule
      \textbf{Level} & \textbf{Task family} & \textbf{Specific tasks} & \textbf{Example question template} \\
      \midrule
      \endhead
      \multirow{7}{*}{$\mathcal{L}_1$}
      & Numeracy & Count objects & \textit{``How many independent sound sources are audible in the clip?''} \\*
      \cmidrule(l){2-4}
      \noalign{\penalty10000}
      & \multirow{2}{=}{Semantics} & Identify event class & \textit{``Which of the following sounds is present in the audio?''} \\*
      & & Identify speech transcript & \textit{``What specific phrase was spoken in the audio?''} \\*
      \cmidrule(l){2-4}
      \noalign{\penalty10000}
      & \multirow{4}{=}{Spatial physics} & Localize azimuth & \textit{``What is the horizontal direction of the event?''} \\*
      & & Localize elevation & \textit{``Is the event coming from above, below, or eye level?''} \\*
      & & Estimate distance & \textit{``Approximately how many meters away is the event?''} \\*
      & & Environment acoustics & \textit{``Which description best matches the room size and reverberation?''} \\
      \midrule
      \multirow{5}{*}{$\mathcal{L}_2$}
      & Binding & Attribute verification & \textit{``Which statement accurately describes the content, location, and distance?''} \\*
      \cmidrule(l){2-4}
      \noalign{\penalty10000}
      & \multirow{2}{=}{Retrieval} & Query location given ID & \textit{``Where is the named event located?''} \\*
      & & Query ID given location & \textit{``What sound is coming from a specified direction and distance?''} \\*
      \cmidrule(l){2-4}
      \noalign{\penalty10000}
      & \multirow{2}{=}{Comparison} & Relative distance & \textit{``Which sound is closer to the listener?''} \\*
      & & Relative direction & \textit{``Is one event positioned to the left or right of another?''} \\
      \noalign{\penalty-10000}
      \midrule
      \multirow{6}{*}{$\mathcal{L}_3$}
      & \multirow{2}{=}{Abduction} & Scene abduction & \textit{``Based on the sounds and acoustics, what is the most likely scene?''} \\*
      & & Causal intent & \textit{``What activity or intent is best implied by the sounds?''} \\*
      \cmidrule(l){2-4}
      \noalign{\penalty10000}
      & \multirow{2}{=}{Counterfactual} & Rotate observer ($180^\circ$) & \textit{``If you turn around $180^\circ$, where would the event be located now?''} \\*
      & & Source removal & \textit{``If one sound is removed, which statement remains true?''} \\*
      \cmidrule(l){2-4}
      \noalign{\penalty10000}
      & \multirow{2}{=}{Logic \& physics} & Multi-hop query & \textit{``Identify the farthest sound source and state its direction.''} \\*
      & & Physical consistency & \textit{``Which statement is physically impossible given the room dimensions?''} \\
      \bottomrule
\end{longtable}
}

\FloatBarrier
\subsection{Input Perturbation Controls}
\label{app:control_perturbations}

We run three controls on the \blue{1,000-question manuscript evaluation} with the final TWNM-SAPO policy. The question-only control replaces every waveform with four-channel silence while keeping the original question and options. The audio-mismatch control deranges audio-question pairs, so every item receives a valid FOA recording from another benchmark scene. The W-only control preserves the original FOA $W$ channel and zeros the directional $X/Y/Z$ channels, retaining non-directional acoustic content while removing first-order directional evidence. \blue{This inference-time perturbation differs from the trained $W$-only control in Appendix~\ref{app:matched_controls}.}

We report these controls with a strict pick-letter audit using Gemini 3 Flash. This differs from semantic-answer judging: the judge is constrained to identify the selected option letter from the model output rather than deciding whether a free-form answer semantically matches the gold option text. We use this audit because corrupted-input outputs often contain repeated options or malformed text, where simple exact parsing can be unstable and semantic-answer judging can over-credit copied option text. Table~\ref{tab:asa_input_controls} therefore reports only the perturbed controls, while the full-FOA exact-MCQA result remains in the main tables. All three controls remain above the 25\% random baseline, which is expected for a four-choice benchmark with text, answer-option, and non-directional acoustic priors. Their scores are also tightly clustered, from 40.4\% to 43.1\%, supporting the main interpretation that the residual control performance comes from priors and degraded non-directional evidence rather than intact spatial grounding.

\begin{table}[!htbp]
  \centering
  \caption{Input-perturbation controls for TWNM-SAPO on the 1,000-question ASA benchmark. Rows use strict pick-letter auditing to parse the selected option from corrupted-input outputs. Values are percentages.}
  \label{tab:asa_input_controls}
  {\small
    \begin{tabular}{@{}llcccc@{}}
      \toprule
      \textbf{Run} & \textbf{Audio input} & $\mathcal{L}_1$ & $\mathcal{L}_2$ & $\mathcal{L}_3$ & \textbf{Overall} \\
      \midrule
      Question-only & Four-channel silence & \textbf{43.12} & 34.05 & 50.60 & \textbf{43.10} \\
      Audio-mismatch & Deranged FOA & 35.32 & \textbf{34.77} & 50.89 & 40.40 \\
      W-only & $W$ kept, $X/Y/Z$ zeroed & 38.96 & 30.47 & \textbf{54.46} & 41.80 \\
      \bottomrule
    \end{tabular}%
  }
\end{table}

Table~\ref{tab:asa_l3_control_priors} breaks out the $\mathcal{L}_3$ control behavior. Observer rotation and physical consistency remain relatively high even with corrupted audio, indicating answer-option and scene-prior structure in those families. Multi-hop and source-removal controls are lower, and all three control variants remain close to one another. Comparing these controls with the full-FOA SAPO row in Table~\ref{tab:main_task_family_results}, the \blue{descriptive differences} over the strongest corrupted control are 28.57 points for causal intent, 28.57 for source removal, 5.35 for observer rotation, 32.14 for multi-hop, 26.79 for physical consistency, and 25.00 for scene abduction. \blue{These differences combine input-distribution and scoring changes and do not isolate the contribution of audio.}

\begin{table}[!htbp]
  \centering
  \caption{Selected $\mathcal{L}_3$ input-perturbation controls under strict pick-letter auditing. Values are percentages.}
  \label{tab:asa_l3_control_priors}
  {\small
    \begin{tabular}{@{}lccc@{}}
      \toprule
      \textbf{Task family} & \textbf{Question-only} & \textbf{Audio-mismatch} & \textbf{W-only} \\
      \midrule
      Causal intent & \textbf{53.57} & \textbf{53.57} & 51.79 \\
      Source removal & 37.50 & 35.71 & \textbf{50.00} \\
      Rotate $180^\circ$ & \textbf{64.29} & \textbf{64.29} & \textbf{64.29} \\
      Multi-hop & \textbf{39.29} & \textbf{39.29} & 35.71 \\
      Physical consistency & \textbf{62.50} & 60.71 & \textbf{62.50} \\
      Scene abduction & 46.43 & 51.79 & \textbf{62.50} \\
      \bottomrule
    \end{tabular}%
  }
\end{table}

\FloatBarrier
\subsection{Full ASA Benchmark Results}
\label{app:full_asa_results}

\blue{Table~\ref{tab:main_task_family_results} reports the complete task-family breakdown in the main text.} Task families are nearly balanced, with 55 or 56 questions each; BAT-SFT is reported from the direct-MCQA semantic-answer audit described in Appendix~\ref{app:bat_diagnostics}, while TWNM stages are scored by exact MCQA matching on the final 1,000-question benchmark.

\FloatBarrier
\subsection{Complete Baseline Audits and Additional LALM Results}
\label{app:additional_baselines}

Table~\ref{tab:additional_lalm_baselines} records the scoring audit behind the compact main table. We use semantic-answer auditing for open-ended outputs because several baselines produce free-form answers rather than a clean option letter, and exact option parsing can undercount correct responses that paraphrase the option text. Exact option extraction is also useful for option-order sensitivity analysis. Because the two audits can differ by several points, the main comparison states the audit protocol for each row and treats cross-audit rows as diagnostic references. For Qwen2.5-Omni, whose outputs were reliably parseable under the exact-option protocol, the main table uses the average of original-order and answer-shuffled exact-option audits; the semantic-judge row documents the alternative audit protocol.

These rows should be read as a baseline audit rather than as a strict same-input leaderboard. \blue{Sci-Phi and SPUR provide complementary FOA-based evidence under their own evaluation protocols (Appendix~\ref{app:spatial_system_comparison}).} BAT is \blue{the spatial-audio reference evaluated in this table} because it consumes spatialized audio, but its interface is binaural and its native tasks do not target hierarchical multi-source ASA reasoning. The monaural API and open-weight rows answer a complementary diagnostic question: how far current scaled audio-language systems can get from semantic, answer-option, and non-directional acoustic evidence alone, and where they degrade once directional binding is required.

The external API baselines are evaluated on the mono FOA W-channel waveform resampled to the service-compatible audio rate. All baselines receive the same audio-only four-choice instruction and no RTSD or metadata. The reported audit uses the dated service identifiers shown in the tables, including GPT-audio 2025-08-28, Qwen3-Omni-Flash 2025-12-01, Step-Audio-R1.1, and Gemini-3.1-Pro Preview, accessed through their provider APIs during the May 2026 audit run. For semantic-answer audits, we use Gemini 3 Flash as the judge model to compare the generated answer with the question, options, and gold option text; the judge never sees the audio or hidden RTSD. No manual repair is applied.

\begin{table}[htbp]
  \centering
  \caption{Additional baseline scoring audits on the 1,000-question ASA benchmark. Values are percentages; the first column names whether each row uses exact option parsing, semantic-answer judging, exact order averaging, or exact MCQA scoring.}
  \label{tab:additional_lalm_baselines}
  {\small
    \begin{tabular}{@{}llcccc@{}}
      \toprule
      \textbf{Model / audit} & \textbf{Input} & $\mathcal{L}_1$ (\%) & $\mathcal{L}_2$ (\%) & $\mathcal{L}_3$ (\%) & \textbf{Overall (\%)} \\
      \midrule
      Qwen2-Audio-7B, exact option & Mono & 41.04 & 31.18 & 30.95 & 34.90 \\
      Qwen2-Audio-7B, semantic judge & Mono & 52.40 & 42.40 & 48.80 & 48.40 \\
      Qwen2.5-Omni-7B, semantic judge & Mono & 49.90 & 41.90 & 51.00 & 48.00 \\
      Qwen2.5-Omni-7B, exact default order & Mono & 61.30 & 47.67 & 68.45 & 59.90 \\
      Qwen2.5-Omni-7B, exact shuffled order & Mono & 55.58 & 51.61 & 65.48 & 57.80 \\
      Qwen2.5-Omni-7B, exact order average & Mono & 58.44 & 49.64 & 66.97 & 58.85 \\
      Qwen3-Omni-Flash, exact parse & Mono & 58.70 & 47.31 & 65.48 & 57.80 \\
      GPT-audio, exact parse & Mono & 52.47 & 55.91 & 67.56 & 58.50 \\
      Step-Audio-R1.1, exact parse & Mono & 59.22 & 48.03 & 71.13 & 60.10 \\
      Gemini-3.1-Pro Preview, exact parse & Mono & 57.92 & 54.12 & 74.11 & 62.30 \\
      TWNM-SAPO, exact MCQA & FOA & \textbf{63.64} & \textbf{69.89} & \textbf{79.76} & \textbf{70.80} \\
      \bottomrule
    \end{tabular}%
  }
\end{table}

\FloatBarrier
\subsection{BAT Adaptation Analysis}
\label{app:bat_diagnostics}

BAT is a spatial audio-language baseline with a different native data format and answer protocol. We report two BAT analyses. First, a binaural-finetuned BAT checkpoint is evaluated on the ASA benchmark as direct MCQA. Under the semantic-answer audit used for the BAT row in the main comparison, with all 1,000 benchmark audios successfully loaded, BAT-SFT obtains $252/1000=25.2\%$ overall. An independent pick-letter audit gives $257/1000=25.7\%$, showing near-random performance under direct ASA prompting. The result is consistent with the training mismatch: BAT's native supervision uses single-source spatialized audio, whereas ASA $\mathcal{L}_2$ questions are all multi-source relation or binding tasks. A single-source inductive bias can therefore map the question to the wrong object or relation with high confidence, producing below-random $\mathcal{L}_2$ behavior when the option set contrasts multiple simultaneous sources.

Second, we evaluate an MCQ-format adapted BAT run after removing the 1,000 benchmark IDs from the adaptation data and balancing answer positions. This run changes the answer-format training target and is therefore treated as a diagnostic rather than as the main BAT baseline. Table~\ref{tab:bat_mcq_diagnostic} reports selected spatial task families where the adapted model remains limited. This pattern suggests that answer-format adaptation improves compliance, whereas the single-source-to-multi-source spatial binding mismatch remains the main bottleneck.

\begin{table}[H]
  \centering
  \caption{Spatial task-pattern audit for MCQ-format adapted BAT on the answer-shuffled benchmark. Selected task families highlight remaining spatial perception, binding, comparison, counterfactual, and multi-hop weaknesses after answer-format adaptation.}
  \label{tab:bat_mcq_diagnostic}
  {\small
    \begin{tabular}{@{}lcc@{}}
      \toprule
      \textbf{Task family} & \textbf{Type} & \textbf{Accuracy (\%)} \\
      \midrule
      Localize Azimuth Direction & Spatial perception & 50.91 \\
      Bind Identity, Location, Distance & Spatial binding & 41.07 \\
      Compare Object Distances & Spatial comparison & 42.86 \\
      Relative Direction Between Objects & Spatial comparison & 38.18 \\
      Counterfactual Rotate $180^\circ$ & Spatial counterfactual & \textbf{51.79} \\
      Multi-hop Composed Query & Spatial reasoning & 46.43 \\
      \bottomrule
    \end{tabular}%
  }
\end{table}

\Needspace{14\baselineskip}
\section{Ablations and Internal Diagnostics}
\label{app:ablation_details}

This appendix collects internal design ablations for TWNM. The projector and attractor ablation tests whether the spatial-semantic interface is necessary before the final policy stage, while the SAPO reward ablation probes which reward terms matter under a reduced rollout budget.

\subsection{Projector and Attractor Ablations}

\blue{Table~\ref{tab:projector_ablation} evaluates projector design before the final SAPO stage. The dense hybrid scores 52.1\% overall, compared with 39.2\% for a single MLP, 46.4\% for the dual tower, 44.5\% without the attractor, and 43.9\% without the attractor encoder. Four parallel spatial transforms process the shared dense representation. Their outputs, together with the semantic and combined views, are preserved before compression.}

\begin{table}[!htbp]
  \centering
  \caption{Projector and attractor ablations on the ASA benchmark. Level columns are semantic-judge accuracy percentages; the overall column counts all 1,000 benchmark items.}
  \label{tab:projector_ablation}
  {\small
    \begin{tabular}{@{}lcccc@{}}
      \toprule
      \textbf{Variant} & $\mathcal{L}_1$ (\%) & $\mathcal{L}_2$ (\%) & $\mathcal{L}_3$ (\%) & \textbf{Overall (\%)} \\
      \midrule
      P0: Single MLP & 37.76 & 27.11 & 51.64 & 39.20 \\
      P1: Dual Tower & 44.88 & 34.55 & \textbf{59.10} & 46.40 \\
      P2: Dense Hybrid & \textbf{56.43} & \textbf{41.88} & 56.55 & \textbf{52.10} \\
      P2 No-Attractor & 45.09 & 36.23 & 52.08 & 44.50 \\
      P2 No-Attractor-Encoder & 42.55 & 39.33 & 54.80 & 43.90 \\
      \bottomrule
    \end{tabular}%
  }
\end{table}

\subsection{SAPO Reward Ablations}

\blue{Table~\ref{tab:sapo_reward_ablation} reports SAPO reward ablations under resource-limited reduced rollouts with semantic-answer scoring. All variants use the same reduced-budget setting. Full SAPO scores 54.0\%; no-format 53.6\%; correctness-only 51.0\%; no-evidence/reference 50.5\%; and no-correctness 46.2\%. The complete reward exceeds correctness-only by 3.0 points and no-evidence/reference by 3.5. Stage~4 updates decoder LoRA while keeping both encoders and the projector frozen.}

\begin{table}[!htbp]
  \centering
  \caption{Reduced-budget SAPO reward ablations. Level columns are semantic-judge accuracy percentages; overall accuracy counts all rollout items.}
  \label{tab:sapo_reward_ablation}
  {\small
    \begin{tabular}{@{}lcccc@{}}
      \toprule
      \textbf{Reward setting} & $\mathcal{L}_1$ (\%) & $\mathcal{L}_2$ (\%) & $\mathcal{L}_3$ (\%) & \textbf{Overall (\%)} \\
      \midrule
      Full SAPO (reduced) & 56.10 & 50.00 & \textbf{66.10} & \textbf{54.00} \\
      No format reward & \textbf{61.40} & 44.70 & 59.60 & 53.60 \\
      Correctness-only reward & 43.20 & 41.00 & 52.90 & 51.00 \\
      No evidence/reference reward & 56.80 & 46.20 & 50.00 & 50.50 \\
      No correctness reward & 58.10 & \textbf{55.30} & 47.20 & 46.20 \\
      \bottomrule
    \end{tabular}%
  }
\end{table}

\Needspace{10\baselineskip}
\section{STARSS23 Real-Recording Diagnostics}
\label{app:starss_real_recording}

This appendix groups the real-recording diagnostics based on STARSS23. The first part analyzes how track-wise SELD outputs behave when converted into source-count tokens. The second part reports the fixed end-to-end STARSS23 QA audit and its audio-perturbation check.

\subsection{SELD Threshold Analysis}
\label{app:seld_threshold}

The encoder comparison in Section~\ref{sec:encoder_analysis} uses source-count, direction, and distance metrics matched to the acoustic-token interface. Table~\ref{tab:einv2_threshold} adds a threshold-sensitivity analysis for converting track-wise SELD outputs into blind source-counting tokens for a language model. EINV2 predicts activity separately for multiple tracks, so a fixed activation threshold can strongly bias the estimated number of active sources.

\begin{table}[H]
  \centering
  \caption{EINV2 source-count accuracy under different activity thresholds. Values are percentages; the evaluation set contains 113 one-source, 61 two-source, and 26 three-source examples.}
  \label{tab:einv2_threshold}
  {\small
    \begin{tabular}{@{}ccccc@{}}
      \toprule
      \textbf{Threshold} & \textbf{1-src (\%, N=113)} & \textbf{2-src (\%, N=61)} & \textbf{3-src (\%, N=26)} & \textbf{Overall (\%)} \\
      \midrule
      0.10 & 0.00 & \textbf{100.00} & \textbf{0.00} & 30.50 \\
      0.45 & 90.30 & 41.00 & \textbf{0.00} & \textbf{63.50} \\
      0.75 & \textbf{100.00} & 0.00 & \textbf{0.00} & 56.50 \\
      \bottomrule
    \end{tabular}%
  }
\end{table}

This analysis complements the localization metrics in Table~\ref{tab:encoder_diagnostics}. It shows that the track activation interface is brittle for downstream tokenization: low thresholds over-activate tracks, high thresholds collapse multi-source scenes into one active source, and three-source counting remains especially fragile.

\subsection{End-to-End STARSS23 QA Details}
\label{app:starss_details}

The STARSS23 QA experiments use real FOA recordings and questions derived from the available spatiotemporal annotations. The fixed 500-question audit contains $210$ $\mathcal{L}_1$ questions, $168$ $\mathcal{L}_2$ questions, and $122$ $\mathcal{L}_3$ questions, generated from 168 STARSS23 recordings and 5,020 candidate clips before filtering. The selected v5 audit items cover fold3 and fold4 recordings and are not used in the LALM SFT or SAPO stages. Because STARSS23 recordings also support Stage-1 sim-to-real adaptation, this is a corpus-level real-recording diagnostic rather than a recording-disjoint held-out benchmark. Table~\ref{tab:starss_detail_runs} lists the runs that completed the fixed audit protocol. The BAT-SFT row covers the successfully loaded subset; all other rows cover the full 500-question STARSS23 QA audit. The perturbation rows keep the same questions but replace audio with Gaussian noise; the resulting drop is used as a coarse audio-perturbation sensitivity check. All entries use attempted-question denominators, so invalid outputs are counted as incorrect.

STARSS23 generated-answer scoring uses the same semantic-answer audit principle as the main baseline audits. Each item is a four-choice question, but models may generate free-form text rather than a clean option letter. The judge receives the question, options, gold option text, and model answer, and outputs a strict JSON correctness verdict; it does not receive the audio, STARSS23 annotation frames, RTSD, sample ID-derived metadata, or the model input waveform. Empty or invalid judge outputs are counted as incorrect after retry handling.

\begin{table}[!htbp]
  \centering
  \caption{Additional STARSS23 real-recording QA results. Values are generated-answer audit percentages over attempted questions. Full-set rows use $N_{\mathcal{L}_1}=210$, $N_{\mathcal{L}_2}=168$, and $N_{\mathcal{L}_3}=122$; BAT-SFT uses the successfully loaded subset with $N_{\mathcal{L}_1}=205$, $N_{\mathcal{L}_2}=139$, and $N_{\mathcal{L}_3}=120$, where the corresponding correct counts are 72, 38, and 40.}
  \label{tab:starss_detail_runs}
  {\small
    \begin{tabular}{@{}lccccc@{}}
      \toprule
      \textbf{Run} & \textbf{$N$} & $\mathcal{L}_1$ (\%) & $\mathcal{L}_2$ (\%) & $\mathcal{L}_3$ (\%) & \textbf{Overall (\%)} \\
      \midrule
      BAT-SFT & 464 & 35.12 & 27.34 & 33.33 & 32.33 \\
      Qwen2.5-Omni full & 500 & 41.90 & \textbf{40.48} & 36.89 & 40.20 \\
      TWNM-SFT & 500 & 41.90 & 38.10 & 47.54 & 42.00 \\
      TWNM-SAPO full & 500 & \textbf{45.24} & 38.69 & \textbf{49.18} & \textbf{44.00} \\
      TWNM-SFT, Gaussian noise & 500 & 22.38 & 17.86 & 31.15 & 23.00 \\
      TWNM-SAPO, Gaussian noise & 500 & 21.90 & 25.60 & 32.79 & 25.80 \\
      \bottomrule
    \end{tabular}%
  }
\end{table}

\Needspace{4\baselineskip}
For a finer perturbation check, we compare one task family under real audio and Gaussian noise. On STARSS23 multi-hop composed queries, TWNM-SAPO reaches $46.34\%$ with real audio but only $14.63\%$ after replacing the waveform with Gaussian noise, showing that this task family is sensitive to the acoustic input.

\Needspace{5\baselineskip}
\blue{\section{Matched Training Controls}
\label{app:matched_controls}}

\begin{table}[!htbp]
  \centering
  \blue{\caption{Matched controls reported in the author response on the 1,000-question ASA benchmark. All rows use exact-option scoring; differences report paired-bootstrap 95\% confidence intervals.}
  \label{tab:matched_controls}}
  \blue{\small
  \begin{tabular}{@{}lrc@{}}
    \toprule
    \textbf{Model / input} & \textbf{Accuracy (\%)} & \textbf{Difference [95\% CI] (pp)} \\
    \midrule
    Question-only & 28.2 & -- \\
    Trained $W$-only, SFT & 49.3 & -- \\
    Full FOA, SFT & 62.0 & $+12.7\ [9.2,16.1]$ vs. $W$-only \\
    Full FOA, SAPO & 70.8 & $+8.8\ [5.6,12.1]$ vs. SFT \\
    \bottomrule
  \end{tabular}}
\end{table}

\blue{\paragraph{Trained $W$-only control.}
The trained $W$-only variant provides a matched control for the complete spatial pathway. From Stage~2 onward, the spatial encoder, projection, and normalization are absent. Stage~2/3 examples, frozen Whisper and AF3/Qwen2.5 bases, P2 interface, optimizer schedule, global batch size, update count, decoding, and exact-option scorer match Full-FOA SFT.}

\blue{Both models stop after SFT, before SAPO. The inference-time $W$-only perturbation in Appendix~\ref{app:control_perturbations} applies the FOA-trained model to audio with directional channels zeroed at test time.}

\blue{\paragraph{Paired results.}
The \href{https://openreview.net/forum?id=51hFveUNno&noteId=NxclRBEoMq}{author response} reports trained $W$-only accuracy of 49.3\% overall and 51.67/40.33/54.25\% at $\mathcal{L}_1/\mathcal{L}_2/\mathcal{L}_3$. Full-FOA SFT scores 62.0\% overall and 72.00/45.67/66.75\% in this matched comparison. The paired overall gain is 12.7 percentage points (paired-bootstrap 95\% CI $[9.2,16.1]$; exact McNemar $p=1.70\times10^{-12}$). Full SFT alone answers 227 questions correctly; $W$-only alone answers 100. Question-only scores 28.2\%. Full-FOA SAPO scores 70.8\%, a paired gain of 8.8 points over SFT (95\% CI $[5.6,12.1]$; $p=2.40\times10^{-7}$). The matched per-level denominators are not specified in the author response. Table~\ref{tab:main_diagnostics} separately reports the submitted stage trajectory. The main-text matched comparison uses the overall paired results.}

\Needspace{5\baselineskip}
\blue{\section{Human Validation}
\label{app:human_validation}}

\blue{Two annotators independently assessed 180 frozen benchmark questions, with 10 items from each of the 18 task families, before seeing the hidden gold answers. Predeclared disagreement or possible-defect criteria triggered independent adjudication of 36 items. Table~\ref{tab:human_agreement} reports inter-annotator agreement and Cohen's $\kappa$ for selected answers, question correctness, answer uniqueness, and audio--question consistency.}

\begin{table}[H]
  \centering
  \blue{\caption{Inter-annotator agreement on the 180-question validation subset. Each row reports counts, percentage agreement, and Cohen's $\kappa$.}
  \label{tab:human_agreement}}
  \blue{\begin{tabular}{@{}lccc@{}}
    \toprule
    \textbf{Aspect} & \textbf{Agree / Total} & \textbf{Agreement (\%)} & $\kappa$ \\
    \midrule
    Selected answer & 153/180 & 85.0 & .81 \\
    Question correctness & 162/180 & 90.0 & .65 \\
    Answer uniqueness & 171/180 & 95.0 & .72 \\
    Audio--question consistency & 158/180 & 87.8 & .70 \\
    \bottomrule
  \end{tabular}}
\end{table}

\blue{After adjudication, selected answers agreed with hidden gold answers on 172 of 180 questions (95.6\%; Wilson 95\% CI $[91.5,97.7]$). Final quality flags were invalid questions (1.7\%), ambiguous or absent answers (2.8\%), audio--question inconsistency (1.1\%), major audio artifacts (3.9\%), and unrateable items (0.0\%). These measurements cover the 180-question subset. The \href{https://openreview.net/forum?id=51hFveUNno&noteId=096ZBIfA5n}{author response} reports the validation results.}

\Needspace{5\baselineskip}
\blue{\section{Spatial-Encoder Binding Diagnostic}
\label{app:binding_diagnostic}}

\blue{We evaluate the frozen spatial encoder's slot outputs before semantic fusion on 1,000 held-out full-simulation scenes containing 1,493 source instances. Predicted slots are assigned to ground-truth sources with Hungarian matching using direction, distance, and objectness costs, with the class-matching cost set to zero. This prevents semantic labels from driving the assignment used to assess binding. Of the matched slots, 666/1,493 predict the correct source class (44.61\%; Wilson 95\% CI $[42.10,47.14]$), 90.56\% have DOA error below $20^\circ$, and 41.39\% satisfy both criteria.}

\blue{Across 404 multi-source scenes, each slot's top-1 class is compared with its own spatially assigned source and, separately, with the other sources in that scene. Own-source agreement is 37.90\%; agreement across the other-source comparisons is 24.23\%. The binding advantage is 13.68 percentage points (scene-bootstrap 95\% CI $[10.64,16.77]$). Two-source scenes show a 13.17-point advantage (95\% CI $[9.37,17.14]$); three-source scenes show 10.67 points (95\% CI $[5.99,15.54]$).}

\blue{Restoring the class-matching cost changes only 4 of 1,493 assignments and leaves class accuracy at 44.61\%. These results show source-specific semantic information associated with spatial slots before fusion. The \href{https://openreview.net/forum?id=51hFveUNno&noteId=096ZBIfA5n}{author response} reports the diagnostic.}

\blue{The evaluated checkpoint uses $K=3$ source slots. Increasing $K$ requires corresponding source-count supervision and encoder training; performance at larger capacities has not been evaluated.}

\Needspace{5\baselineskip}
\blue{\section{Comparison with Spatial Audio-Language Systems}
\label{app:spatial_system_comparison}}

\blue{Sci-Phi evaluates spatial features through feature-zeroing ablations \citep{jiang2026sciphi}. In Table~4 of arXiv:2510.05542v1, results averaged over one to four sources with synthetic room impulse responses show that retaining spatial and spectral features yields 91.5\% source-count accuracy and 82.9\% direction accuracy, compared with 25.0\% and 9.1\% when spatial features are zeroed. Direction accuracy measures classification among 26 three-dimensional regions. TupleScore, the geometric mean of What, Where, and When scores, increases from 0.562 to 0.783. These rows evaluate feature availability within a trained model. TWNM's trained $W$-only comparison removes the spatial branch throughout training and obtains 49.3\% versus 62.0\% for full-FOA SFT under matched training and exact-MCQA scoring.}

\blue{SPUR augments Audio Flamingo 3 with an FOA spatial encoder, adapter, LoRA, and SPUR-Set training \citep{sakshi2025spur}. Its Table~2 reports increases in Spatial Consistency from 4.94 to 7.50, Reasoning Depth from 3.63 to 5.51, and Relevance from 5.83 to 8.24, each scored out of 10 by an LLM judge. Error Rate decreases from 48.48\% to 22.99\%; it counts responses with contradictions, omissions, hallucinations, or vagueness. This comparison jointly changes spatial input, model components, and spatial-QA supervision. The three studies provide complementary within-study evidence under distinct datasets, tasks, and metrics; their numerical values do not define a cross-study ranking.}

\newpage
\blue{
\renewcommand{\answerYes}[1][]{[Yes]#1}
\renewcommand{\answerNo}[1][]{[No]#1}
\renewcommand{\answerNA}[1][]{[N/A]#1}
\section*{NeurIPS Paper Checklist}

\begin{enumerate}

\Needspace{9\baselineskip}
\item {\bf Claims}
    \item[] Question: Do the main claims made in the abstract and introduction accurately reflect the paper's contributions and scope?
    \item[] Answer: \answerYes{}
    \item[] Justification: The abstract and introduction state the paper's intended contributions: spatial audio simulation, FOA-conditioned model design, progressive training, and ASA benchmark evaluation. Appendix~\ref{app:limitations} states the current scope and limitations.
    \item[] Guidelines:
    \begin{itemize}
        \item The answer \answerNA{} means that the abstract and introduction do not include the claims made in the paper.
        \item The abstract and/or introduction should clearly state the claims made, including the contributions made in the paper and important assumptions and limitations. A \answerNo{} or \answerNA{} answer to this question will not be perceived well by the reviewers.
        \item The claims made should match theoretical and experimental results, and reflect how much the results can be expected to generalize to other settings.
        \item It is fine to include aspirational goals as motivation as long as it is clear that these goals are not attained by the paper.
    \end{itemize}

\Needspace{9\baselineskip}
\item {\bf Limitations}
    \item[] Question: Does the paper discuss the limitations of the work performed by the authors?
    \item[] Answer: \answerYes{}
    \item[] Justification: Appendix~\ref{app:limitations} discusses the simulation-heavy evaluation setting, the limited real-world end-to-end validation, multichannel input assumptions, and potential privacy-sensitive uses of spatial audio localization.
    \item[] Guidelines:
    \begin{itemize}
        \item The answer \answerNA{} means that the paper has no limitation while the answer \answerNo{} means that the paper has limitations, but those are not discussed in the paper.
        \item The authors are encouraged to create a separate ``Limitations'' section in their paper.
        \item The paper should point out any strong assumptions and how robust the results are to violations of these assumptions (e.g., independence assumptions, noiseless settings, model well-specification, asymptotic approximations only holding locally). The authors should reflect on how these assumptions might be violated in practice and what the implications would be.
        \item The authors should reflect on the scope of the claims made, e.g., if the approach was only tested on a few datasets or with a few runs. In general, empirical results often depend on implicit assumptions, which should be articulated.
        \item The authors should reflect on the factors that influence the performance of the approach. For example, a facial recognition algorithm may perform poorly when image resolution is low or images are taken in low lighting. Or a speech-to-text system might not be used reliably to provide closed captions for online lectures because it fails to handle technical jargon.
        \item The authors should discuss the computational efficiency of the proposed algorithms and how they scale with dataset size.
        \item If applicable, the authors should discuss possible limitations of their approach to address problems of privacy and fairness.
        \item While the authors might fear that complete honesty about limitations might be used by reviewers as grounds for rejection, a worse outcome might be that reviewers discover limitations that aren't acknowledged in the paper. The authors should use their best judgment and recognize that individual actions in favor of transparency play an important role in developing norms that preserve the integrity of the community. Reviewers will be specifically instructed to not penalize honesty concerning limitations.
    \end{itemize}

\Needspace{9\baselineskip}
\item {\bf Theory assumptions and proofs}
    \item[] Question: For each theoretical result, does the paper provide the full set of assumptions and a complete (and correct) proof?
    \item[] Answer: \answerNA{}
    \item[] Justification: The paper is empirical and does not claim new theoretical results requiring formal proofs.
    \item[] Guidelines:
    \begin{itemize}
        \item The answer \answerNA{} means that the paper does not include theoretical results.
        \item All the theorems, formulas, and proofs in the paper should be numbered and cross-referenced.
        \item All assumptions should be clearly stated or referenced in the statement of any theorems.
        \item The proofs can either appear in the main paper or the supplemental material, but if they appear in the supplemental material, the authors are encouraged to provide a short proof sketch to provide intuition.
        \item Inversely, any informal proof provided in the core of the paper should be complemented by formal proofs provided in appendix or supplemental material.
        \item Theorems and Lemmas that the proof relies upon should be properly referenced.
    \end{itemize}

\Needspace{13\baselineskip}
\item {\bf Experimental result reproducibility}
    \item[] Question: Does the paper fully disclose all the information needed to reproduce the main experimental results of the paper to the extent that it affects the main claims and/or conclusions of the paper (regardless of whether the code and data are provided or not)?
    \item[] Answer: \answerNo{}
    \item[] Justification: The main paper and appendices disclose the model architecture, simulation procedure, benchmark construction, training curriculum, reward design, scoring rules, hyperparameters, and compute scale needed to understand and audit the reported experiments. The anonymized supplement provides a compact architecture prototype with core model code and a representative architecture configuration. It does not provide a complete turnkey rerun of all paper-scale training jobs, closed-API baselines, evaluation audits, full corpus rendering, or large checkpoint artifacts.
    \item[] Guidelines:
    \begin{itemize}
        \item The answer \answerNA{} means that the paper does not include experiments.
        \item If the paper includes experiments, a \answerNo{} answer to this question will not be perceived well by the reviewers: Making the paper reproducible is important, regardless of whether the code and data are provided or not.
        \item If the contribution is a dataset and\slash or model, the authors should describe the steps taken to make their results reproducible or verifiable.
        \item Depending on the contribution, reproducibility can be accomplished in various ways. For example, if the contribution is a novel architecture, describing the architecture fully might suffice, or if the contribution is a specific model and empirical evaluation, it may be necessary to either make it possible for others to replicate the model with the same dataset, or provide access to the model. In general. releasing code and data is often one good way to accomplish this, but reproducibility can also be provided via detailed instructions for how to replicate the results, access to a hosted model (e.g., in the case of a large language model), releasing of a model checkpoint, or other means that are appropriate to the research performed.
        \item While NeurIPS does not require releasing code, the conference does require all submissions to provide some reasonable avenue for reproducibility, which may depend on the nature of the contribution. For example
        \begin{enumerate}
            \item If the contribution is primarily a new algorithm, the paper should make it clear how to reproduce that algorithm.
            \item If the contribution is primarily a new model architecture, the paper should describe the architecture clearly and fully.
            \item If the contribution is a new model (e.g., a large language model), then there should either be a way to access this model for reproducing the results or a way to reproduce the model (e.g., with an open-source dataset or instructions for how to construct the dataset).
            \item We recognize that reproducibility may be tricky in some cases, in which case authors are welcome to describe the particular way they provide for reproducibility. In the case of closed-source models, it may be that access to the model is limited in some way (e.g., to registered users), but it should be possible for other researchers to have some path to reproducing or verifying the results.
        \end{enumerate}
    \end{itemize}

\Needspace{12\baselineskip}
\item {\bf Open access to data and code}
    \item[] Question: Does the paper provide open access to the data and code, with sufficient instructions to faithfully reproduce the main experimental results, as described in supplemental material?
    \item[] Answer: \answerYes{}
    \item[] Justification: The submission includes an anonymized architecture-prototype code package with the core TWNM model implementation and a representative architecture configuration. The package is code-facing rather than data-complete: it does not include the full rendered corpora, closed-model API access, evaluation-audit machinery, large checkpoints, or cluster-specific storage needed to faithfully rerun every main result from scratch.
    \item[] Guidelines:
    \begin{itemize}
        \item The answer \answerNA{} means that paper does not include experiments requiring code.
        \item Please see the NeurIPS code and data submission guidelines (\url{https://neurips.cc/public/guides/CodeSubmissionPolicy}) for more details.
        \item While we encourage the release of code and data, we understand that this might not be possible, so \answerNo{} is an acceptable answer. Papers cannot be rejected simply for not including code, unless this is central to the contribution (e.g., for a new open-source benchmark).
        \item The instructions should contain the exact command and environment needed to run to reproduce the results. See the NeurIPS code and data submission guidelines (\url{https://neurips.cc/public/guides/CodeSubmissionPolicy}) for more details.
        \item The authors should provide instructions on data access and preparation, including how to access the raw data, preprocessed data, intermediate data, and generated data, etc.
        \item The authors should provide scripts to reproduce all experimental results for the new proposed method and baselines. If only a subset of experiments are reproducible, they should state which ones are omitted from the script and why.
        \item At submission time, to preserve anonymity, the authors should release anonymized versions (if applicable).
        \item Providing as much information as possible in supplemental material (appended to the paper) is recommended, but including URLs to data and code is permitted.
    \end{itemize}

\Needspace{11\baselineskip}
\item {\bf Experimental setting/details}
    \item[] Question: Does the paper specify all the training and test details (e.g., data splits, hyperparameters, how they were chosen, type of optimizer) necessary to understand the results?
    \item[] Answer: \answerYes{}
    \item[] Justification: Sections~\ref{sec:method} and~\ref{sec:exp}, together with Appendix sections on data simulation and training implementation, specify the architecture, data generation procedure, optimizer, batch sizes, learning rates, LoRA configuration, loss terms, scoring rules, and SAPO reward setup. Appendix~\ref{app:matched_controls} specifies the matched training and evaluation conditions reported in the author response.
    \item[] Guidelines:
    \begin{itemize}
        \item The answer \answerNA{} means that the paper does not include experiments.
        \item The experimental setting should be presented in the core of the paper to a level of detail that is necessary to appreciate the results and make sense of them.
        \item The full details can be provided either with the code, in appendix, or as supplemental material.
    \end{itemize}

\Needspace{12\baselineskip}
\item {\bf Experiment statistical significance}
    \item[] Question: Does the paper report error bars suitably and correctly defined or other appropriate information about the statistical significance of the experiments?
    \item[] Answer: \answerYes{}
    \item[] Justification: Appendix~\ref{app:audit_protocols} defines the Wilson 95\% binomial interval for the primary ASA comparison. Appendix~\ref{app:matched_controls} reports paired bootstrap intervals and exact McNemar tests for the matched controls, Appendix~\ref{app:human_validation} reports a Wilson interval for human validation, and Appendix~\ref{app:binding_diagnostic} reports scene-bootstrap intervals for the binding diagnostic. These intervals quantify evaluation-sample uncertainty under the stated protocols.
    \item[] Guidelines:
    \begin{itemize}
        \item The answer \answerNA{} means that the paper does not include experiments.
        \item The authors should answer \answerYes{} if the results are accompanied by error bars, confidence intervals, or statistical significance tests, at least for the experiments that support the main claims of the paper.
        \item The factors of variability that the error bars are capturing should be clearly stated (for example, train/test split, initialization, random drawing of some parameter, or overall run with given experimental conditions).
        \item The method for calculating the error bars should be explained (closed form formula, call to a library function, bootstrap, etc.)
        \item The assumptions made should be given (e.g., Normally distributed errors).
        \item It should be clear whether the error bar is the standard deviation or the standard error of the mean.
        \item It is OK to report 1-sigma error bars, but one should state it. The authors should preferably report a 2-sigma error bar than state that they have a 96\% CI, if the hypothesis of Normality of errors is not verified.
        \item For asymmetric distributions, the authors should be careful not to show in tables or figures symmetric error bars that would yield results that are out of range (e.g., negative error rates).
        \item If error bars are reported in tables or plots, the authors should explain in the text how they were calculated and reference the corresponding figures or tables in the text.
    \end{itemize}

\Needspace{10\baselineskip}
\item {\bf Experiments compute resources}
    \item[] Question: For each experiment, does the paper provide sufficient information on the computer resources (type of compute workers, memory, time of execution) needed to reproduce the experiments?
    \item[] Answer: \answerYes{}
    \item[] Justification: Appendix~\ref{app:training_details} reports the optimization and implementation settings used for the reported model variants. Detailed wall-clock and cluster-run accounting is not central to the paper's claims and is not tabulated.
    \item[] Guidelines:
    \begin{itemize}
        \item The answer \answerNA{} means that the paper does not include experiments.
        \item The paper should indicate the type of compute workers CPU or GPU, internal cluster, or cloud provider, including relevant memory and storage.
        \item The paper should provide the amount of compute required for each of the individual experimental runs as well as estimate the total compute.
        \item The paper should disclose whether the full research project required more compute than the experiments reported in the paper (e.g., preliminary or failed experiments that didn't make it into the paper).
    \end{itemize}

\Needspace{10\baselineskip}
\item {\bf Code of ethics}
    \item[] Question: Does the research conducted in the paper conform, in every respect, with the NeurIPS Code of Ethics \url{https://neurips.cc/public/EthicsGuidelines}?
    \item[] Answer: \answerYes{}
    \item[] Justification: The work uses public or simulated audio data, does not involve private recordings or human-subject experiments, and includes a discussion of privacy-sensitive spatial localization risks in Appendix~\ref{app:limitations}.
    \item[] Guidelines:
    \begin{itemize}
        \item The answer \answerNA{} means that the authors have not reviewed the NeurIPS Code of Ethics.
        \item If the authors answer \answerNo, they should explain the special circumstances that require a deviation from the Code of Ethics.
        \item The authors should make sure to preserve anonymity (e.g., if there is a special consideration due to laws or regulations in their jurisdiction).
    \end{itemize}

\Needspace{10\baselineskip}
\item {\bf Broader impacts}
    \item[] Question: Does the paper discuss both potential positive societal impacts and negative societal impacts of the work performed?
    \item[] Answer: \answerYes{}
    \item[] Justification: Appendix~\ref{app:limitations} discusses positive uses in robotics, augmented reality, assistive listening, and spatial scene understanding, as well as negative uses such as acoustic surveillance and privacy-invasive inference.
    \item[] Guidelines:
    \begin{itemize}
        \item The answer \answerNA{} means that there is no societal impact of the work performed.
        \item If the authors answer \answerNA{} or \answerNo, they should explain why their work has no societal impact or why the paper does not address societal impact.
        \item Examples of negative societal impacts include potential malicious or unintended uses (e.g., disinformation, generating fake profiles, surveillance), fairness considerations (e.g., deployment of technologies that could make decisions that unfairly impact specific groups), privacy considerations, and security considerations.
        \item The conference expects that many papers will be foundational research and not tied to particular applications, let alone deployments. However, if there is a direct path to any negative applications, the authors should point it out. For example, it is legitimate to point out that an improvement in the quality of generative models could be used to generate Deepfakes for disinformation. On the other hand, it is not needed to point out that a generic algorithm for optimizing neural networks could enable people to train models that generate Deepfakes faster.
        \item The authors should consider possible harms that could arise when the technology is being used as intended and functioning correctly, harms that could arise when the technology is being used as intended but gives incorrect results, and harms following from (intentional or unintentional) misuse of the technology.
        \item If there are negative societal impacts, the authors could also discuss possible mitigation strategies (e.g., gated release of models, providing defenses in addition to attacks, mechanisms for monitoring misuse, mechanisms to monitor how a system learns from feedback over time, improving the efficiency and accessibility of ML).
    \end{itemize}

\Needspace{11\baselineskip}
\item {\bf Safeguards}
    \item[] Question: Does the paper describe safeguards that have been put in place for responsible release of data or models that have a high risk for misuse (e.g., pre-trained language models, image generators, or scraped datasets)?
    \item[] Answer: \answerNA{}
    \item[] Justification: The submission does not release a deployment-ready surveillance system, scraped personal dataset, or high-risk model weights. The paper and appendices document research-use scope and provenance for the benchmark assets, while the anonymized supplement contains only a compact architecture prototype.
    \item[] Guidelines:
    \begin{itemize}
        \item The answer \answerNA{} means that the paper poses no such risks.
        \item Released models that have a high risk for misuse or dual-use should be released with necessary safeguards to allow for controlled use of the model, for example by requiring that users adhere to usage guidelines or restrictions to access the model or implementing safety filters.
        \item Datasets that have been scraped from the Internet could pose safety risks. The authors should describe how they avoided releasing unsafe images.
        \item We recognize that providing effective safeguards is challenging, and many papers do not require this, but we encourage authors to take this into account and make a best faith effort.
    \end{itemize}

\Needspace{11\baselineskip}
\item {\bf Licenses for existing assets}
    \item[] Question: Are the creators or original owners of assets (e.g., code, data, models), used in the paper, properly credited and are the license and terms of use explicitly mentioned and properly respected?
    \item[] Answer: \answerYes{}
    \item[] Justification: The paper cites the public datasets, pretrained model families, simulation software, and closed API services used in the work. The anonymized supplement contains only a compact architecture prototype and does not relicense upstream raw assets.
    \item[] Guidelines:
    \begin{itemize}
        \item The answer \answerNA{} means that the paper does not use existing assets.
        \item The authors should cite the original paper that produced the code package or dataset.
        \item The authors should state which version of the asset is used and, if possible, include a URL.
        \item The name of the license (e.g., CC-BY 4.0) should be included for each asset.
        \item For scraped data from a particular source (e.g., website), the copyright and terms of service of that source should be provided.
        \item If assets are released, the license, copyright information, and terms of use in the package should be provided. For popular datasets, \url{paperswithcode.com/datasets} has curated licenses for some datasets. Their licensing guide can help determine the license of a dataset.
        \item For existing datasets that are re-packaged, both the original license and the license of the derived asset (if it has changed) should be provided.
        \item If this information is not available online, the authors are encouraged to reach out to the asset's creators.
    \end{itemize}

\Needspace{10\baselineskip}
\item {\bf New assets}
    \item[] Question: Are new assets introduced in the paper well documented and is the documentation provided alongside the assets?
    \item[] Answer: \answerYes{}
    \item[] Justification: The paper documents the benchmark construction pipeline, task hierarchy, simulation parameters, and prompt generation procedure. The anonymized release package includes generated-asset documentation and license metadata.
    \item[] Guidelines:
    \begin{itemize}
        \item The answer \answerNA{} means that the paper does not release new assets.
        \item Researchers should communicate the details of the dataset\slash code\slash model as part of their submissions via structured templates. This includes details about training, license, limitations, etc.
        \item The paper should discuss whether and how consent was obtained from people whose asset is used.
        \item At submission time, remember to anonymize your assets (if applicable). You can either create an anonymized URL or include an anonymized zip file.
    \end{itemize}

\Needspace{10\baselineskip}
\item {\bf Crowdsourcing and research with human subjects}
    \item[] Question: For crowdsourcing experiments and research with human subjects, does the paper include the full text of instructions given to participants and screenshots, if applicable, as well as details about compensation (if any)?
    \item[] Answer: \answerNA{}
    \item[] Justification: The reported experiments do not involve crowdsourcing or research with human subjects. The manual review of 180 benchmark questions in Appendix~\ref{app:human_validation} is benchmark quality control.
    \item[] Guidelines:
    \begin{itemize}
        \item The answer \answerNA{} means that the paper does not involve crowdsourcing nor research with human subjects.
        \item Including this information in the supplemental material is fine, but if the main contribution of the paper involves human subjects, then as much detail as possible should be included in the main paper.
        \item According to the NeurIPS Code of Ethics, workers involved in data collection, curation, or other labor should be paid at least the minimum wage in the country of the data collector.
    \end{itemize}

\Needspace{10\baselineskip}
\item {\bf Institutional review board (IRB) approvals or equivalent for research with human subjects}
    \item[] Question: Does the paper describe potential risks incurred by study participants, whether such risks were disclosed to the subjects, and whether Institutional Review Board (IRB) approvals (or an equivalent approval/review based on the requirements of your country or institution) were obtained?
    \item[] Answer: \answerNA{}
    \item[] Justification: The reported experiments do not involve study participants or human-subject data collection, so IRB approval is not applicable.
    \item[] Guidelines:
    \begin{itemize}
        \item The answer \answerNA{} means that the paper does not involve crowdsourcing nor research with human subjects.
        \item Depending on the country in which research is conducted, IRB approval (or equivalent) may be required for any human subjects research. If you obtained IRB approval, you should clearly state this in the paper.
        \item We recognize that the procedures for this may vary significantly between institutions and locations, and we expect authors to adhere to the NeurIPS Code of Ethics and the guidelines for their institution.
        \item For initial submissions, do not include any information that would break anonymity (if applicable), such as the institution conducting the review.
    \end{itemize}

\Needspace{10\baselineskip}
\item {\bf Declaration of LLM usage}
    \item[] Question: Does the paper describe the usage of LLMs if it is an important, original, or non-standard component of the core methods in this research? Note that if the LLM is used only for writing, editing, or formatting purposes and does \emph{not} impact the core methodology, scientific rigor, or originality of the research, declaration is not required.
    \item[] Answer: \answerYes{}
    \item[] Justification: LLMs are core components of the research: the proposed system uses an LLM decoder, and benchmark construction uses a teacher LLM to convert simulator-derived RTSD metadata into QA items. This usage is described in Section~\ref{subsec:datasets} and Appendix~\ref{app:llm_usage}.
    \item[] Guidelines:
    \begin{itemize}
        \item The answer \answerNA{} means that the core method development in this research does not involve LLMs as any important, original, or non-standard components.
        \item Please refer to our LLM policy in the NeurIPS handbook for what should or should not be described.
    \end{itemize}

\end{enumerate}
}

\end{document}